\documentclass[a4paper,fleqn,useAMS,usenatbib]{mnras}

\pdfoutput=1

\makeatletter
\newsavebox\myboxA
\newsavebox\myboxB
\newlength\mylenA

\usepackage{natbib}
\usepackage{mathptmx}

\usepackage[T1]{fontenc}
\usepackage{ae,aecompl}

\usepackage[dvips]{graphicx}
\usepackage{amsmath}
\usepackage{amsfonts}
\usepackage{amssymb}
\usepackage{comment}
\usepackage{bm} 
\usepackage[dvipsnames]{xcolor}
\usepackage[]{hyperref}
        
\hypersetup{colorlinks=true,urlcolor=MidnightBlue,pdfpagelabels=true,hypertexnames=true,plainpages=false,naturalnames=true,pdftitle={The Multiplicity Distribution of Kepler's Exoplanets},pdfauthor={Cool Worlds Lab},linkcolor=WildStrawberry,citecolor=ForestGreen}

\usepackage{tikz}
 
\usepackage{soul}
\usepackage{enumitem}

\newcommand{\pdf}{\mathrm{Pr}}

\title[On planetary systems as ordered sequences]{
On planetary systems as ordered sequences}
\author[Sandford et al.]{Emily Sandford,$^{1,2}$\thanks{E-mail:
\href{mailto:es835@cam.ac.uk}{es835@cam.ac.uk}} David Kipping$^{3,4}$ and Michael Collins$^{5,6}$ \\
$^{1}$Astrophysics Group, Cavendish Laboratory, J.J.Thomson Avenue, Cambridge, CB3 0HE, UK \\
$^{2}$Gonville \& Caius College, Trinity Street, Cambridge, CB2 1TA, UK\\
$^{3}$Dept. of Astronomy, Columbia University, 550 W. 120th Street, New York, NY 10027, USA \\
$^{4}$Center for Computational Astrophysics, Flatiron Institute, 162 5th Avenue, New York, NY 10010, USA\\
$^{5}$Dept. of Computer Science, Columbia University, 1214 Amsterdam Avenue, New York, NY 10027, USA \\
$^{6}$Data Science Institute, Columbia University, 550 W. 120th Street, New York, NY 10027, USA
}

\date{Accepted . Received ; in original form }

\pubyear{2020}

\begin{document}
\label{firstpage}
\pagerange{\pageref{firstpage}--\pageref{lastpage}}
\maketitle

\begin{abstract}
A planetary system consists of a host star and one or more planets, arranged into a particular configuration. Here, we consider what information belongs to the configuration, or ordering, of 4286 \textit{Kepler} planets in their 3277 planetary systems. First, we train a neural network model to predict the radius and period of a planet based on the properties of its host star and the radii and period of its neighbors. The mean absolute error of the predictions of the trained model is a factor of 2.1 better than the MAE of the predictions of a naive model which draws randomly from dynamically allowable periods and radii. Second, we adapt a model used for unsupervised part-of-speech tagging in computational linguistics to investigate whether planets or planetary systems fall into natural categories with physically interpretable ``grammatical rules.'' The model identifies two robust groups of planetary systems: (1) compact multi-planet systems and (2) systems around giant stars ($\log{g} \lesssim 4.0$), although the latter group is strongly sculpted by the selection bias of the transit method. These results reinforce the idea that planetary systems are not random sequences---instead, as a population, they contain predictable patterns that can provide insight into the formation and evolution of planetary systems.

\end{abstract}


\section{Introduction}
\label{sec:chap6_intro}

The planets belonging to a planetary system are not expected to be random or independent. We broadly expect them to have formed from the same protoplanetary disk, collapsed initially from the same cloud, around the same star (see \citealt{williams11} for a review). We expect early-forming planets to determine which planets form later. Within our own Solar System, for example, Jupiter and Saturn are thought to have formed early, migrated inward, and truncated the protoplanetary disk at $\sim 1$ AU, which left relatively little mass to form the terrestrial planets over the subsequent $\sim 50$ Myr \citep{walsh11,batygin15}. Within exoplanetary systems, we observe correlations between the properties of sibling planets, suggesting similar interdependence \citep{ciardi13,millholland17,weiss:2018}. 

After the disk dissipates, we expect the planets to continue interacting dynamically, through orbital migrations, resonances, and scatterings; the present-day configuration of the system, from its multiplicity \citep{nesvorny11,sandford19}, to the spacing and ordering of the planets (\citealt{lissauer11pop,fabrycky14,weiss:2018,kipping18entropy}; see \citealt{winn15} for a review), to their mutual inclinations and consequent co-transiting probability \citep{tremaine12,fang:2012,figueira:2012,weissbein:2012,ballard:2016}, will naturally depend on this dynamical history. 

Thus, the planetary system---the star and its planets, in their specific configuration---encodes its formation and dynamical history; there is information in the individual components, but also in their \textit{arrangement}. It is therefore valuable, especially now that we know of so many exoplanetary systems, to, in the words of \cite{gilbert20}, ``treat the \textit{system} as the fundamental unit [of exoplanet science]" and investigate planets within the context of their siblings and host stars.

Historically, the relationship of planets to their context has been difficult to investigate because of the combinatoric proliferation of relationships between planets as system multiplicity grows, and because of the lack of an obvious way to compare systems of different multiplicity. \cite{gilbert20} address these problems by defining seven scalar statistics, each of which captures some aspect of the planetary system overall: three relevant examples are the system multiplicity; the monotonicity, which describes how strictly size-ordered the planets are; and the characteristic spacing of the planets in mutual Hill radii. 

Here, we take a different approach to studying planetary systems, inspired by the study of linguistics. Linguistics concerns itself not just with vocabularies of individual words, but also with their arrangement into meaningful sequences; a particular arrangement of words conveys information which the same words, randomized, do not. Furthermore, any particular language has grammatical rules and conventions that govern these arrangements, which is interesting for two reasons: first, because grammatical logic allows us to make inferences about missing words, and second, because studying the set of allowable arrangements as a whole can allow us to infer the underlying rules. In natural language processing, these tasks are known respectively as ``masked language modeling'' (e.g. \citealt{devlin2018}) and ``grammar induction'' (see \citealt{grammarInductionReview} for a recent review).

By analogy, we concern ourselves here with planetary systems as ordered sequences, the structure of which is governed, presumably self-consistently, by the rules of physics and probability. We adapt models from computational linguistics \citep{stratos19, mcallester18} to explore two questions. First, can we infer the properties of a missing or unobserved planet based on its observed context? If yes, these predictions would be useful in planet searches, to narrow the range of parameter space in which we might expect to find the planet. Second, what can the patterns of planets in systems tell us? Can we infer anything about the underlying ``grammatical structure'' of planetary systems, if there is such a structure?

In Section~\ref{sec:chap6_data}, we describe our input catalog of KOIs arranged in systems. In Section~\ref{sec:chap6_regression}, we investigate the regression question: how well can we predict the period and radius of an exoplanet based on its context, i.e. the properties of its host star and the periods and radii of its neighbour planets? Furthermore, which contextual information is most important in making those predictions? In Section~\ref{sec:chap6_classification}, we turn to classification, and apply a model based on linguistic part-of-speech tagging to explore the categories of exoplanet and the ``grammatical roles" they play in their systems. In Section~\ref{sec:chap6_discussion}, we discuss the results.

\section{Input catalog}
\label{sec:chap6_data}
The goal of this work is to model the relationships between planets and their surrounding system context. For this investigation, we limit ourselves to planets discovered by the \textit{Kepler} mission. We do not expect \textit{Kepler} planetary systems to be complete, because of the inherent biases of the transit method and the detection efficiency of the \textit{Kepler} pipeline (see e.g. \citealt{zink19}), but we nevertheless expect the arrangement of these systems to contain some interesting information.

To construct our input catalog, we downloaded the list of \textit{Kepler} DR25 Objects of Interest (KOIs) from the NASA Exoplanet Archive (NEA; \citealt{akeson:2013}).\footnote{Accessed 30 September 2020.} We selected planetary systems containing only KOIs with ``Disposition Using Kepler Data" of ``CANDIDATE" and cross-matched these  with the catalog of \cite{chen:2018}, who predicted masses for the DR25 KOIs using \texttt{forecaster} \citep{chen:2017}. For KOIs without a mass prediction from this catalog, we used \texttt{forecaster}'s \texttt{Rstat2M} function to make a mass prediction ourselves, based on the NEA-reported planetary radius. Although we do not use the planet mass directly as an input feature to our models, it is necessary to evaluate the stability of each planetary system.

We also discarded any system which contained a KOI with $R_p > 100\ R_{\oplus}$ (to eliminate unphysically big entries in the catalog) or \texttt{forecaster}-predicted $M_p > 0.08\ M_{\odot}$, the Jovian-star boundary identified by \citealt{chen:2017}. We are implicitly assuming by this cut that brown dwarfs belong in the planet sample---the sample includes 14 objects with \texttt{forecaster}-predicted masses greater than $13\ M_J$.

After these cuts, our catalog comprised 4286 KOIs, grouped into 3277 systems. These systems range in multiplicity from 1 to 7 KOIs each, where the count of each multiplicity is given in Table~\ref{tab:mults}.


We note here that we do not make any effort to account for the selection bias of the transit method in our sample: it is certain that there exist additional planets in these systems which either do not transit or transit at too low a signal-to-noise ratio to be detected. For our purposes, this incompleteness does not matter---if we find that we are able to predict the properties of a planet based on its surrounding context, that will remain true even if additional planets are later discovered in the same system. Likewise, any grammatical structure we find in these incomplete systems may be incomplete, but will not be rendered incorrect by further planet discoveries.

We decided to limit our investigation to four salient features of each planetary system: the effective temperature $T_\mathrm{eff}$ [K] and surface gravity $\log_{10}{(g\  [\mathrm{cm}/\mathrm{s}^2])}$ of its host star (hereafter simply $\log{g}$, as it is denoted in the NEA), and the radius $R_p$ [$R_\oplus$] and period $P$ [days] of each planet. We choose $R_p$ and $P$ because they are simply interpretable, physically meaningful quantities. Period (which has the additional advantage of being directly measurable from transit light curves) encodes the ordering of planets outward from their star, as well as some information about their insolation. Planetary radius---although it is not directly observable from transits, and depends on isochrone modeling of the host star---contains information about both composition (e.g., whether the planet has a thick atmosphere) and equilibrium temperature (e.g., how thermally inflated the atmosphere is), and is an axis along which we know there to be meaningful boundaries between categories of planets, such as the small rocky planets and super-Earths on opposite sides of the radius gap at $1.5-2.0R_\oplus$ \citep{fulton:2017}. Likewise, we choose $T_\mathrm{eff}$ and $\log{g}$ because they are simple to interpret, capture a lot of information about the host star, and are directly spectroscopically observable.

\begin{figure*}
\begin{center}
\includegraphics[width=\textwidth]{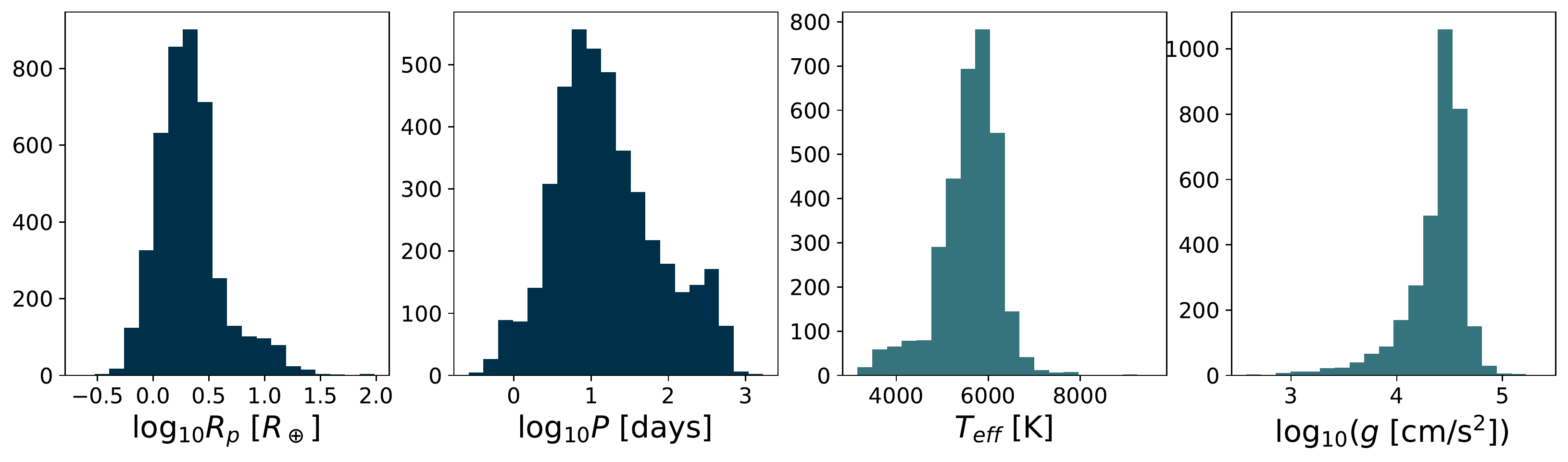}
\caption{Histograms of the two planetary (left) and two stellar (right) features of interest across our catalog of KOIs. The two planetary-feature histograms are over the 4286 planets, and the stellar-feature histograms are over the 3277 systems.}
\label{fig:features}
\end{center}
\end{figure*}

Because $R_p$ and $P$ span 2 and 3 orders of magnitude, respectively, we take $\log_{10}$ of both. Histograms of these four features are presented in Figure~\ref{fig:features}.

\begin{table}
\caption{Multiplicities in the final subset of 4286 KOIs, grouped into 3277 systems, considered in this work.
Taking the sum of each multiplicity by its count yields 4286, as expected.} 
\centering 
\begin{tabular}{c c c c c} 
\hline\hline 
Multiplicity, $m$ & Count \\ [0.5ex] 
\hline 
1 & 2601 \\
2 & 448 \\
3 & 151 \\
4 & 53 \\
5 & 21 \\
6 & 2 \\
7 & 1 \\
[1ex]
\hline\hline 
\end{tabular}
\label{tab:mults} 
\end{table}

\section{Regression: Prediction of planet properties from system context}
\label{sec:chap6_regression}
The first question we ask of our KOI catalog is: what can we say about the properties of a \textit{target} planet, based only on its \textit{context}? We define \textit{context} as consisting of the host star's  $T_\mathrm{eff}$ and $\log{g}$, as well as the radii and periods of the other planets in the system. We seek to predict the target planet's $R_p$ and $P$ based on this contextual information. 

We note again that we do not attempt to correct for the selection bias of the transit method or the incompleteness of the \textit{Kepler} catalog in making these predictions, for two reasons. Firstly, if we find that the system context contains useful information in predicting the target planet properties, that will be true even if the system is incomplete, i.e. even if subsequent planets are later discovered in the same system. Second, our training and test data sets both consist of \textit{Kepler} systems, subject to exactly the same selection effects, and we will not be attempting to extrapolate our conclusions to data that is differently biased.

If these predictions are accurate, this method could be used to search for suspected additional planets in non-dynamically-packed planetary systems: for example, a strong period prior on a missing planet from contextual information could be used to fold the system's light or RV curves to search for a coherent signal from that planet. 

In Appendix~\ref{sec:appendix}, we investigate this question analytically: if we assume that the planetary system overall is dynamically stable, and we know the periods and radii of two adjacent planets, what constraint can we place on the period and radius of a hypothetical planet between them? We find that the analytic constraints we derive this way are so broad as to be practically irrelevant, so we do not consider them further. 

For a full analytic framework to address this question, see \cite{dietrich20}. This work evaluates the likelihood of an unseen planet of a given period and radius existing in a known planetary system, based on choices of the underlying distributions of period, radius, and inclination across the exoplanet population, and subject to dynamical stability considerations.

\begin{figure*}
\begin{center}
\includegraphics[width=\textwidth]{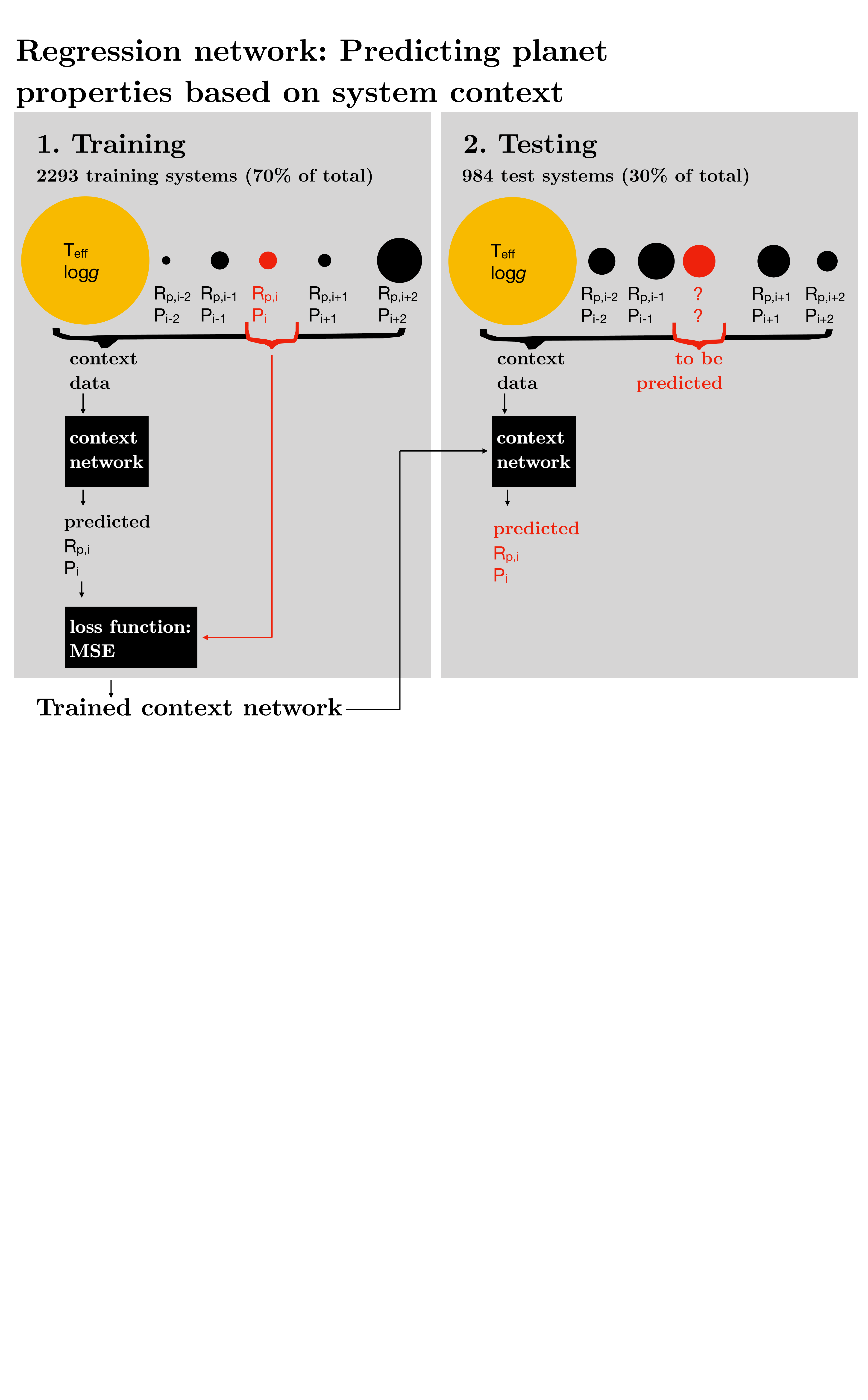}
\caption{A schematic diagram of the machine learning approach to the \textit{regression} problem: predicting the properties of an unobserved middle planet bounded by four observed planets.}
\label{fig:regressionModel}
\end{center}
\end{figure*}

Instead we explore a computational approach to predicting the properties of a planet based on its system context: training a neural network model. A diagram of this approach is presented in Figure~\ref{fig:regressionModel}.

We begin by dividing our data set of 3277 \textit{Kepler} systems into a training set of 2293 systems (70\% of the total) and a test set of 984 systems (the remaining 30\%). 

For each \textit{target planet} in the training set, the network's goal is to take as input the planet's context, and provide as output a prediction of the log planetary radius $\log_{10}{R_p}$ and log period $\log_{10}{P}$, which can then be compared to the true values. 

Specifically, the network's contextual input consists of:
\begin{enumerate}
    \item the stellar $T_\mathrm{eff}$ and $\log{g}$;
    \item a vector of $\log_{10}{P}$ of the $w$ planets immediately inner to the target planet and the $w$ planets immediately outer to the target planet, where $w$ is a network hyperparameter called the \textit{context width};
    \item a corresponding vector of $\log_{10}{R_p}$ of these $2w$ neighbour planets.
\end{enumerate}

In other words, for each target planet, the network sees the planet's host star and the neighbouring planets within a window of width $2w$ centered on the target planet. Of course, because the maximum multiplicity of systems in our data set is $m=7$ and these systems are heavily skewed toward low multiplicity, most planets will not have $w$ inner and $w$ outer neighbours, even for $w=1$. In this case, the period and radius vectors are zero-padded such that they still have length $2w$. Based on testing and considering the tradeoff between capturing more information about high-multiplicity systems (high $w$) and training speed (low $w$), we choose $w = 2$.

Our network is implemented in \texttt{pytorch}. It has two fully-connected hidden layers of width 10 and 5, respectively. For the hidden layers, we use a rectified linear unit (ReLU) activation function; for the output layer, which needs to generate real-valued predictions of arbitrary sign, we use a linear activation function. 

This architecture is a simplified version of the classification network described below in section~\ref{sec:chap6_classification} (which came chronologically first and was itself inspired by the architecture employed by \cite{stratos19}). We initially used the same architecture for the regression task as for the classification task, where the two hidden layers have 100 and 10 neurons respectively, but found after some experimentation that we could reduce the number of neurons in the network by approximately an order of magnitude while still achieving an equally good cost on the training set, substantially improving training speed.

We train the network with an \texttt{Adam} optimizer \citep{kingma14} with learning rate $0.002$. (This is the value used by \cite{stratos19}, although we experimented with values between 0.001 and 0.02 and found that training speed and ultimate network performance were insensitive to the learning rate.) We train the network in batches of 100 training planets over 500 epochs, which (by visual inspection of cost vs. epoch number) is more than sufficient for the network cost to decline to a roughly constant value. To all neurons, we impose a dropout probability of 1\% at every training epoch to discourage overfitting (although we experimented with values between 1 and 10\% and found that again, the results were not sensitive to this parameter). Using these hyperparameters, 500 training epochs take of order 30 seconds on a 2.9 GHz Intel Core i5 processor.

The cost function the network minimizes over training is the mean squared error:
\begin{equation}
    MSE = \frac{1}{N}\sum_{i=1}^N ||\mathbfit{X}_{i,\mathrm{true}} - \mathbfit{X}_{i,\mathrm{pred}}||^2,
\end{equation}
where $\mathbfit{X}_{i,\mathrm{true}}$ is the vector of \textit{true} values $[\log_{10}{R_{p,\mathrm{target}}},\ \log_{10}{P_\mathrm{target}}]$ for target planet $i$, $\mathbfit{X}_{i,\mathrm{pred}}$ is the corresponding network prediction, and $N$ is the number of planets in the training set (or, in practice, training batch). We note that this cost function does not account for the observational uncertainty in $\log_{10}{R_{p,\mathrm{true}}}$ and  $\log_{10}{P_\mathrm{true}}$: in other words, we are training the network to predict the optimal values of $\log_{10}{R_{p,\mathrm{true}}}$ and $\log_{10}{P_\mathrm{true}}$, not their full credible intervals.

\begin{figure}
\begin{center}
\includegraphics[width=0.45\textwidth]{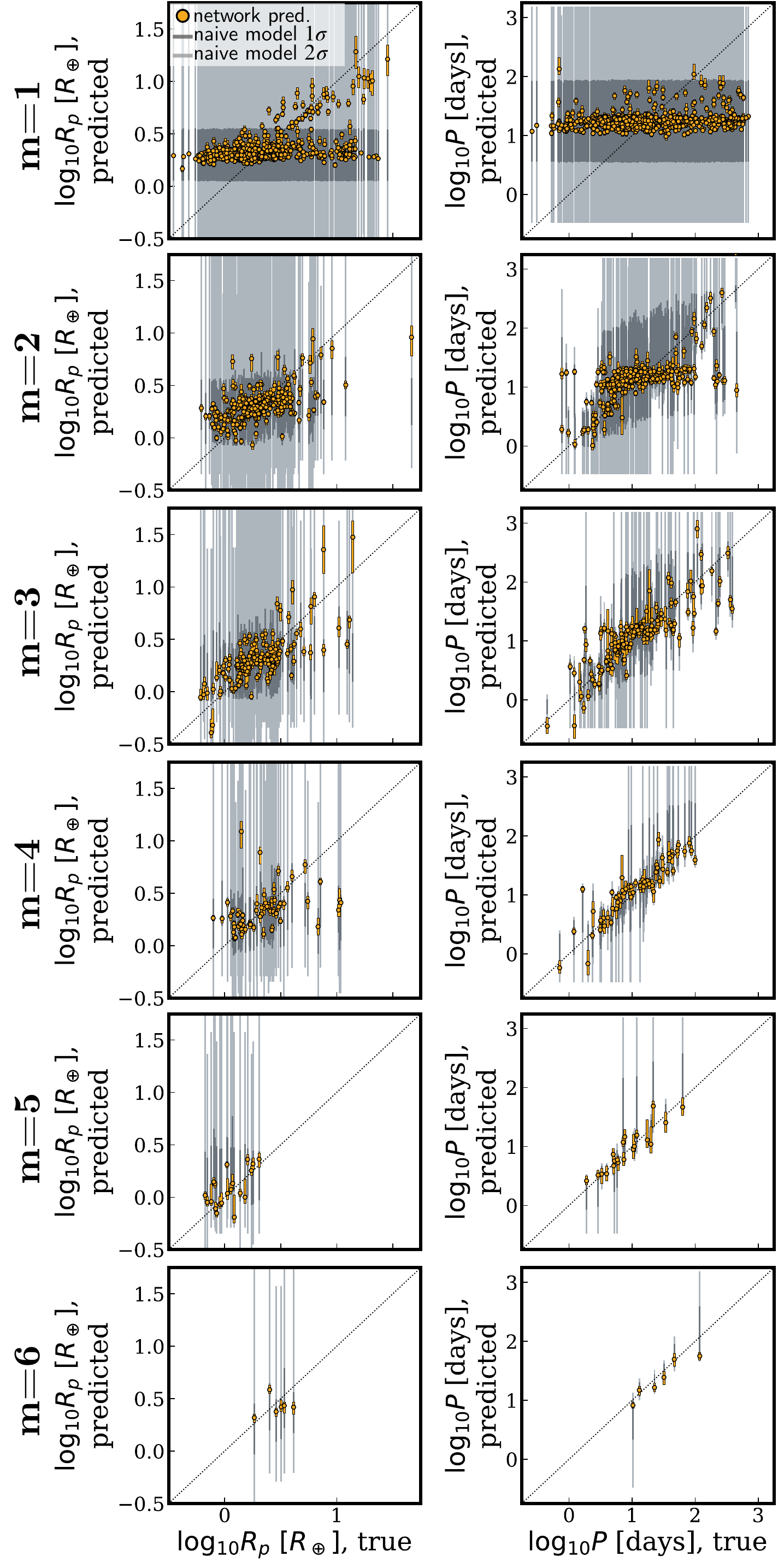}
\caption{Predictions of planet radius (left column) and period (right column) vs. their true values for the 984 planetary systems in the test set, as a function of increasing system multiplicity (rows). In yellow are the network predictions; the yellow circle is the median over the 100 random network initializations, and the yellow bar is the $1\sigma$ confidence interval. In dark (light) gray are the $1\sigma$ ($2\sigma$) confidence intervals of the ``naive" model (see text).}
\label{fig:regressionResults}
\end{center}
\end{figure}

To establish the uncertainty of the network's $\log_{10}{R_{p,\mathrm{true}}}$ and  $\log_{10}{P_\mathrm{true}}$ predictions, we use the ``deep ensembles'' approach recommended by \cite{caldeira20}: we train the network 100 times with different random initializations of the network weights, then let each of these 100 versions make a prediction of $\log_{10}{R_{p}}$ and  $\log_{10}{P}$ for each planet in the test set. In Figure~\ref{fig:regressionResults}, we plot the results of this approach for the 984 test set systems: we plot the median over the 100 predictions for each test set planet as a yellow circle, and the $1\sigma$ confidence interval as a yellow bar.

We also experiment with training the ensemble of 100 networks on a greatly expanded training set that better represents the observational uncertainties of the input data. To construct this training set, we take every planet in the training set and make 100 draws of its contextual data ($T_\mathrm{eff}$, $\log{g}$, $\log_{10}{P}$ of the $2w$ context planets, $\log_{10}{R_p}$ of the $2w$ context planets). Along each dimension, the distribution we draw from is a Gaussian centred on the NEA-reported value, with standard deviation equal to whichever is greater of the NEA-reported upper and lower uncertainties. To train the 100 networks on this hundredfold-expanded training set took substantially longer than on the original training set, so we increased the batch size to 1000 training planets. In the end, however, the results of the network model trained on the expanded training set were not quantitatively different from those on the original training set. We therefore present the results from the original training set, where each training planet is only represented once.

In Figure~\ref{fig:regressionResults}, for comparison to the network model's predictions, we also plot the ``predictions" of an extremely naive model: for each target planet, we ``predict" the radius and period by making 10000 random draws of planets from the training set which satisfy the basic stability criterion $P_\mathrm{inner} < P_\mathrm{draw} < P_\mathrm{outer}$, where $P_\mathrm{inner}$ is the period of the target planet's nearest inner neighbor, and $P_\mathrm{outer}$ is the period of the target planet's nearest outer neighbor. If the target planet is the innermost (outermost) of its system, we adopt as $P_\mathrm{inner}$ ($P_\mathrm{outer}$) the minimum (maximum) period of any planet in our training set. We plot the $1$ and $2\sigma$ confidence intervals over the 10000 draws as gray bars in Figure~\ref{fig:regressionResults}. We note that in the period-prediction scatterplots (right column), you can clearly pick out the planets which are innermost (outermost) in their systems, as these are the planets for which the gray bars extend all the way to the lower (upper) limit of the plotted period range. The gray bars for all of the one-planet systems, for example, fill the entire period range.

As Figure~\ref{fig:regressionResults} shows, the network predictions for both radius and period are no better than random for the vast majority of the 1-planet systems, but improve with system multiplicity. This indicates that the network is learning primarily from the other planets in the system, not from the stellar features.

\begin{figure}
\begin{center}
\includegraphics[width=0.45\textwidth]{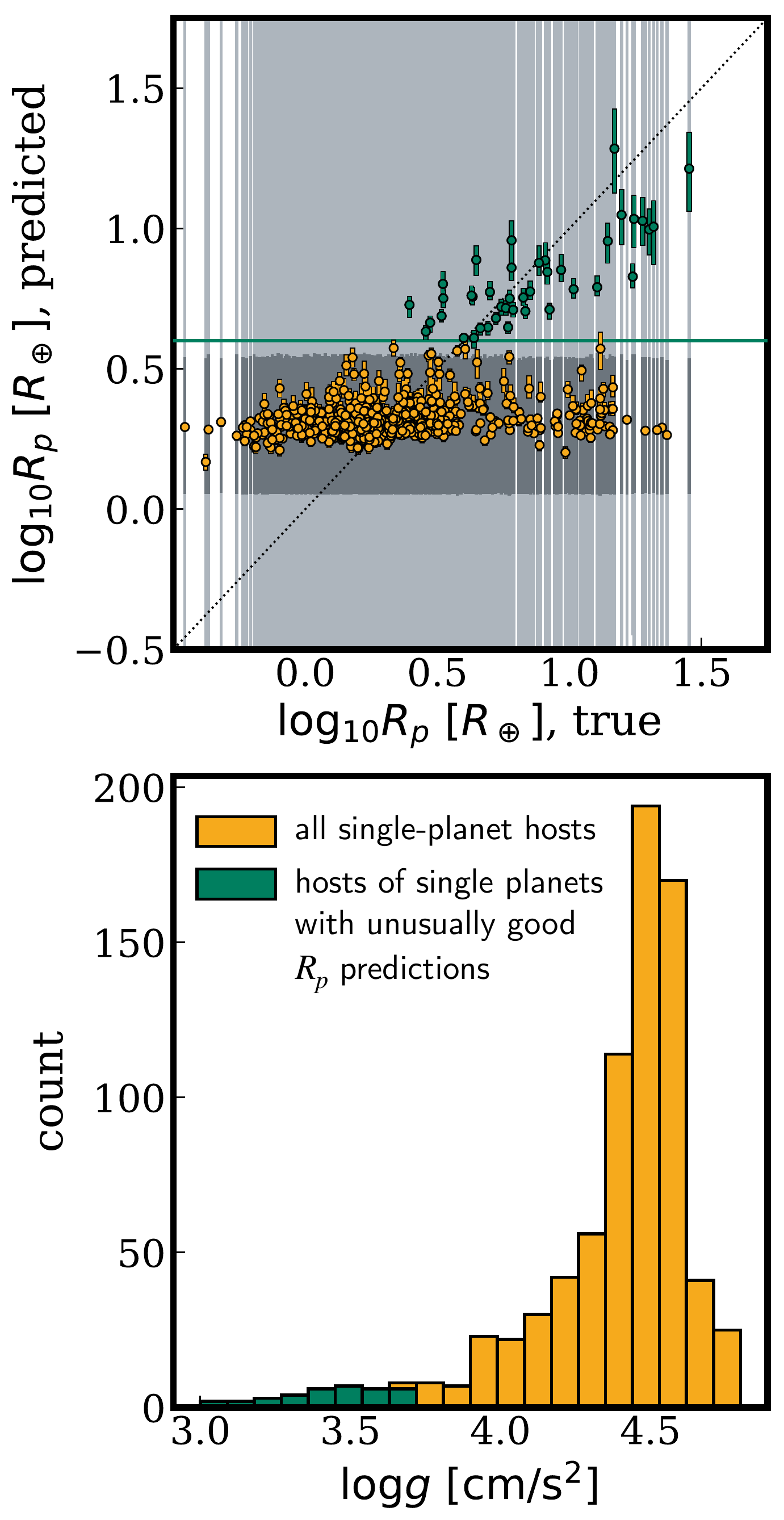}
\caption{Top panel: An enlarged version of the first panel from Figure~\ref{fig:regressionResults}, showing the predicted vs. true planetary radius for the single-planet systems in the test set. In green are the subset of planets with network-predicted $\log_{10}R_p > 0.6$ (horizontal line), which have unusually accurate network radius predictions. Lower panel: A histogram of $\log{g}$ values for the host stars of the single-planet systems in the test set (yellow), showing that the host stars of the planets with unusually good predictions (green) are all giants, with $\log{g} < 3.7$.}
\label{fig:goodRp}
\end{center}
\end{figure}

An exception is a small subset of high radius single-planet systems for which the network is able to make unusually accurate predictions: these planets are highlighted in green in the upper panel of Figure~\ref{fig:goodRp}. (We select these planets as those for which the network predicts $\log_{10}R_p > 0.6$; we plot this selection criterion as a horizontal line in the upper panel of Figure~\ref{fig:goodRp}.)  These are the giant planets orbiting giant stars, as shown in the lower panel of Figure~\ref{fig:goodRp}; the host stars of the single planets with good radius predictions universally have $\log{g} < 3.7$. The network is picking up on the correlation between $R_p$ and $\log{g}$ for planets orbiting giant stars in the KOI data set, the result of selection bias in the transit method: the only planets that can be seen to transit against bright giant stars are giant planets. 


For period, especially, it makes sense that the network learns mainly from the context of neighbouring planets---after all, the network is directly given the bracketing periods of the neighbour planets. However, at multiplicities $m > 2$, the network does better than the naive model for middle planets as well as inner/outermost planets, meaning that the network's improvement over the naive model does not result purely from our choice to bracket the inner/outermost planet periods by $P_\mathrm{minimum}$ and $P_\mathrm{maximum}$ when drawing random planets for the naive model. 

\begin{figure}
\begin{center}
\includegraphics[width=0.45\textwidth]{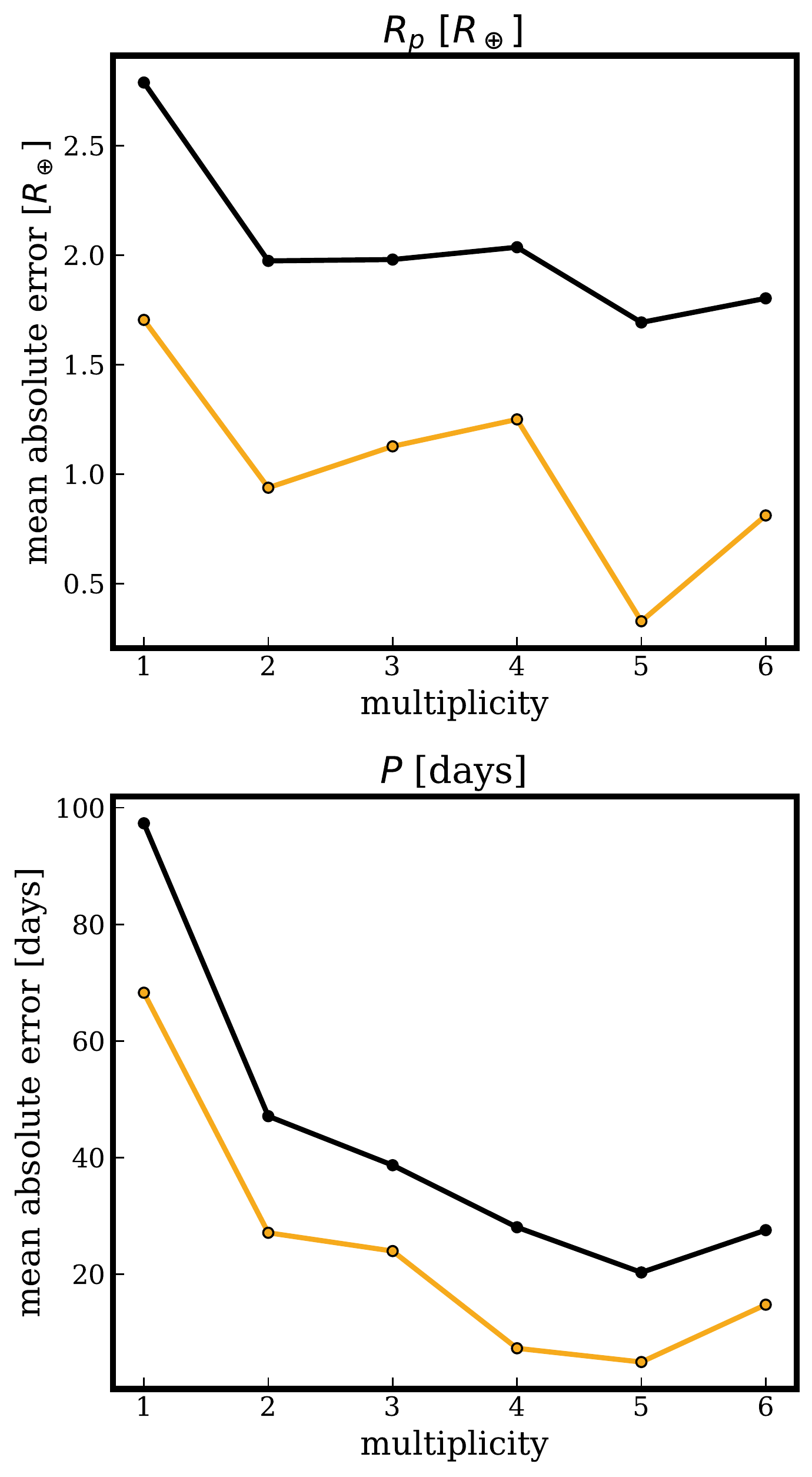}
\caption{The mean absolute error per planet of the network (yellow) and naive model (black) predictions, as a function of multiplicity, in linear (rather than $\log_{10}$) $R_p$ and $P$ space. The network consistently outperforms the naive model: its MAE is smaller by (on average, weighted by the number of planets per multiplicity) a factor of 2.1 in both radius and period.}
\label{fig:MAEcomp_linear}
\end{center}
\end{figure}

A summary of the network vs. naive model performance is presented in Figure~\ref{fig:MAEcomp_linear}, which shows the mean absolute error (MAE) in both radius and period as a function of system multiplicity, transformed into linear $R_p$ and $P$ space, for the trained network (yellow) and the naive model (black). The mean absolute error is defined:
\begin{equation}
    \mathrm{MAE} = \frac{1}{N}\sum_{i=1}^N ||x_{i,\mathrm{true}} - x_{i,\mathrm{pred}}||,
\end{equation}
where N is the number of planets, $x_{i,\mathrm{true}}$ is the true value of $R_p$ (or $P$), and $x_{i,\mathrm{pred}}$ is the predicted value of $R_p$ (or $P$). We choose to plot MAE in linear space to make it easier to judge the accuracy of the two models' predictions in physically meaningful units.

Because the mean absolute error is already normalized by the number of planets, we can compare the different multiplicities in this plot directly. The MAE curves for the network and the naive model trace similar shapes for both radius and period. The radius predictions of both models improve overall with increasing multiplicity, although not monotonically; this is unsurprising given the very small number of systems in the data set with three or more planets. The period predictions of both models become more accurate for multiplicities up to $m = 5$, then worsen for $m=6$; this could once again be the result of the very small number of $m \geq 5$ systems in the data set.

As evidenced by Figure~\ref{fig:MAEcomp_linear}, The network consistently outperforms the naive model. When averaged over all the planets in the test set, the MAE of the naive model's predictions---of both radius and period---is a factor of 2.1 larger than the MAE of the network's predictions. In other words, training the neural network model resulted in a roughly twofold improvement over the naive model in predicting the period and radius of an unseen planet based on its planetary system context.

\section{Classification: The grammar of planets and planetary systems}
\label{sec:chap6_classification}
We next turn to the question of planet grammar. In this section, we adapt a model used for part-of-speech tagging in computational linguistics to ask: do planets fall into natural categories with different roles in their systems, as words fall into parts of speech? If these categories exist, can we deduce them from the set of observed planetary systems? Finally, what can we infer about the underlying rules (or ``grammar'') governing the arrangement of planetary systems? 

We begin by exploring the analogous linguistic question that inspired this work. What is part-of-speech tagging, why is it useful, and how is it done?

\subsection{A model for our model: part-of-speech tagging}\label{subsec:linguistics}

Part-of-speech tagging is the problem of categorizing the words in a body of text by their grammatical function or behavior. In the preceding sentence, for example, ``is" is a verb; ``problem", ``words", ``body", ``text", ``function", and ``behavior" are nouns, etc. This task is deceptively simple: often, the part of speech of a word changes depending on its context, so it has no single true classification. For example, in other sentences, ``words", ``text", and ``function" can all be verbs, as in, ``She words the sentence carefully," ``I will text you later," or ``The computer cannot function." 

Part-of-speech tagging is interesting and useful because it represents sentences abstractly. It captures the patterns of words organized according to a grammar, independent of the choice of words in any particular sentence. For example, ``The bird sings" and ``The plot thickens" both follow the same part-of-speech pattern, ``article noun verb," despite different meaning. The abstraction to ``article noun verb" reveals the common structure. 

Tagging an entire corpus of English text in this way would reveal several common patterns: ``noun verb" if the sentence ends in a period, but ``verb noun" if it ends in a question mark; verbs and adjectives allowed to occur in sequence, but not prepositions. These patterns in turn hint at the underlying grammatical rules of the language. 

Over the past two decades, computational linguists have made great strides in \textit{unsupervised} part-of-speech tagging (see \citealt{christodoulopolous10} for a review), in which a model learns to identify the correct parts of speech without training on a data set labeled with the correct answers. Because grammatical patterns in sentences exist regardless of the content of those sentences, it is not especially necessary to train part-of-speech tagging models on a labeled training set; part-of-speech tagging can be treated more like a clustering problem, where words are grouped together according to similar behavior (e.g. \citealt{brown92}). For our purposes, where we do not have ``correct'' planet labels to train on, unsupervised learning is ideal.

Here, we investigate a particular unsupervised machine learning technique introduced by \cite{stratos19}, based on theoretical work by \cite{mcallester18}, and adapt it to classify planets in systems. This technique, called mutual information maximization, simultaneously trains \textit{two} models to predict the part-of-speech (class membership) of any word (planet). The first model, called the \textit{target network}, takes as input the target word (target planet); the second model, called the \textit{context network}, takes as input the surrounding context words (star and neighbouring planets). Both networks output a prediction of the target's class membership, and the cost function rewards agreement between these predictions. In other words, the goal is to maximize the mutual information---the amount of information learned about one variable by measuring another---between the target network's representation of the target (based on the target itself) and the context network's representation of the target (based on the context).

\subsubsection{The mutual information cost function}

Here, we explore this cost function in slightly more detail. This will be a simplified treatment; for the full derivation and information-theoretic justification, see \cite{stratos19}. 

Following the derivation of \cite{stratos19} section 3.2 (and adopting the same notation), let us label our target planet as $x \in X$, where $X$ is the space of all possible planets, and our planet context as $y \in Y$, where $Y$ is the space of all possible contexts. If we aim to classify the planets into $N_\mathrm{classes}$ classes, our goal is to train the target network to assign a class label $z \in {1,...,N_\mathrm{classes}}$ to the target, to train the context network to assign a class label $z \in {1,...,N_\mathrm{classes}}$ to the target, and for those assigned labels to agree as often as possible.

Let us call the output of the target network $p(z|x)$. This is a vector of probabilities of length $N_\mathrm{classes}$, where e.g. $p(1|x)$ is the probability that planet $x$ belongs to class 1, etc. Similarly, let us call the output of the context network $q(z|y)$.


Let us assume first that we know $q(z|y)$ and we wish to train the target network such that $p(z|x)$ matches it as well as possible. We can reward that agreement by minimizing the cross-entropy $H(q,p)$:
\begin{equation}
    H(q,p) = \mathrm{E} [-\sum_z q(z|y) \ln{p(z|x)}],
\end{equation}

where the expectation value is over all target-context pairs $(x,y)$ in the training set. However, we do not actually know $q(z|y)$, and rewarding agreement alone would allow the two networks to trivially agree by e.g. assigning all planets to the same class. So we must simultaneously penalize that behavior, i.e. reward assigning examples to different classes. 

This can be achieved by maximizing the entropy of $q(z|y)$ over the different choices of label $z$. First we marginalize $q(z|y)$ over all $y$, so we are left with the distribution of labels $q(z)$. Let $Z$ represent a random draw from this distribution. We wish to maximize the entropy of $Z$:
\begin{equation}
    H(Z) = - \sum_z q(z)\ln{q(z)}.
\end{equation}

The overall cost function to be 
\textit{maximized} is then $-1$ times the cross-entropy term, to reward agreement, plus the label entropy term, to penalize trivial agreement. Mathematically, we wish to maximize:
\begin{equation}
    J = H(Z) - H(q,p),
\end{equation}

which \cite{mcallester18} shows to be a lower bound on the mutual information between target $x$ and context $y$.


\subsubsection{Network design}

\begin{figure*}
\begin{center}
\includegraphics[width=\textwidth]{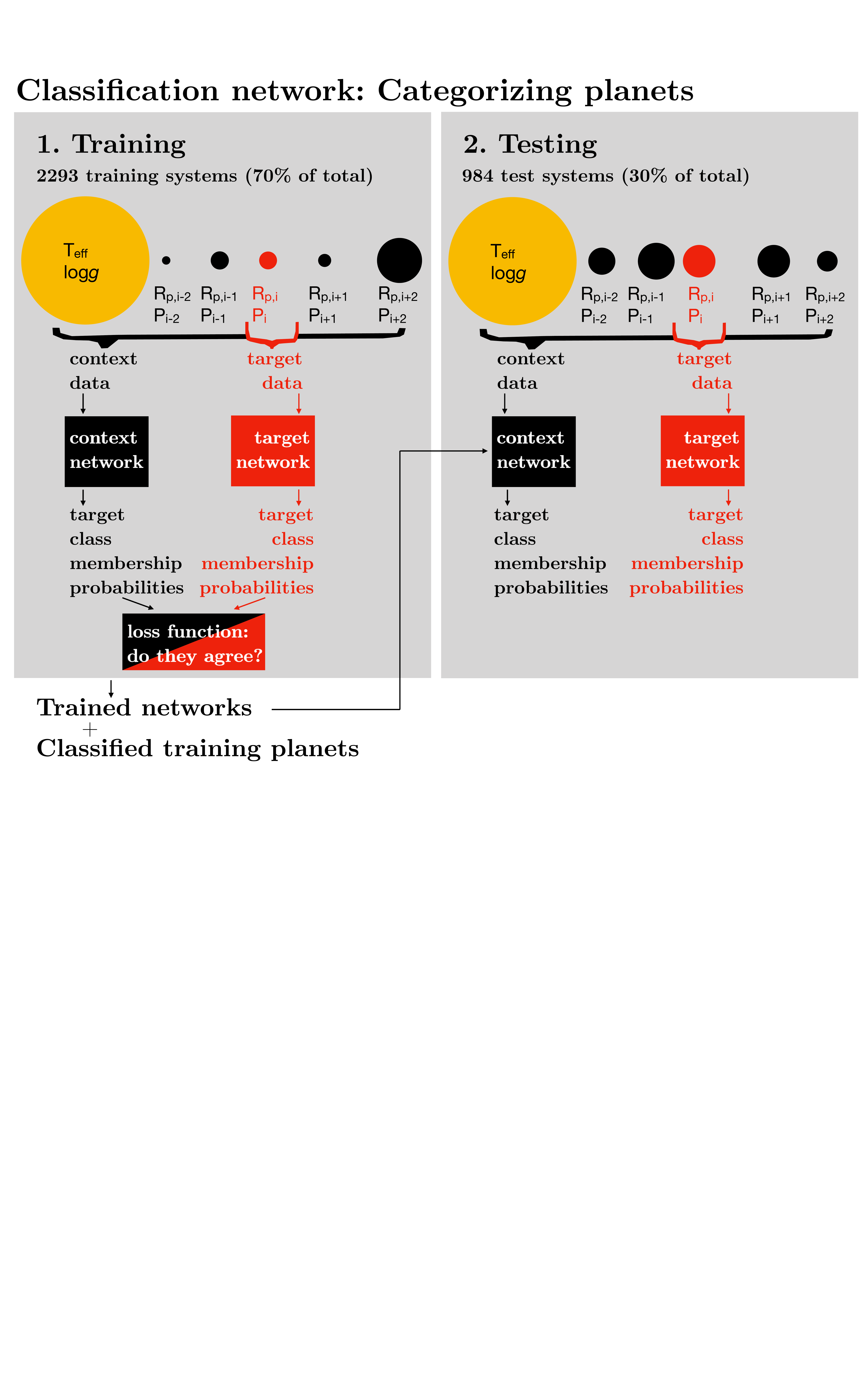}
\caption{A schematic diagram of the machine learning approach to \textit{classifying} planetary systems, which is adapted from the part-of-speech tagger of \protect\cite{stratos19}.}
\label{fig:classificationModel}
\end{center}
\end{figure*}

To apply this mutual information maximization framework to our planets, we first design the target and context networks, then test them on a (highly artificial) data set of simulated planetary systems. A diagram of our classification model is presented in Figure~\ref{fig:classificationModel}.

The target and context networks (again implemented in \texttt{pytorch}) have the same internal architecture, but with different inputs. The target network takes as input the vector $[\log_{10}R_{p,\mathrm{target}}, \log_{10}P_{\mathrm{target}}]$. The context network (as in Section~\ref{sec:chap6_regression}) takes as input the stellar features $T_\mathrm{eff}$ and $\log{g}$, as well as a length-$2w$ vector of the $\log_{10}{R_p}$ of the $w$ inner and $w$ outer context planets, and a corresponding length-$2w$ vector of the $\log_{10}{P}$ of these planets. As in the regression problem, we adopt a context width $w=2$ and zero-pad the context vectors where there are fewer than $2w$ context planets.

Both networks have two fully-connected hidden layers of width 100 and 10, respectively, with ReLU activations. The output layer is of size $N_\mathrm{classes}$, and we apply a softmax function to its output to transform it into a vector of class membership probabilities (i.e. the first entry is the probability that the particular training example belongs to class 1, etc.). $N_\mathrm{classes}$, the number of classes into which the network will attempt to sort the planets, must be chosen before the network can be trained; for our experiments below (see section~\ref{subsec:classified}), we train the network multiple times with different choices of $N_\mathrm{classes}$ and compare the results. 

This architecture was inspired by the context network used for part-of-speech tagging by \cite{stratos19}, which has a single fully-connected hidden layer with of order 1000 neurons (depending on the user's choice of context width). We initially copied this architecture directly, but found that reorganizing the neurons into two hidden layers of size 100 and 10 resulted in much better performance on the test set, without introducing more network weights and therefore without slowing the training speed.

We again train the networks with an \texttt{Adam} optimizer with learning rate 0.002 in batches of 100 training examples over 500 epochs. To all neurons, we again impose a dropout probability of 1\% at every training epoch to discourage overfitting. The classification network, like the regression network, was insensitive to learning rate; however, unlike for the regression network, we found that increasing the dropout probability increased the chances that the classification network would experience a vanishing gradient and fail to converge.

For this choice of hyperparameters, 500 training epochs takes of order 1 minute on a 2.9 GHz Intel Core i5 processor, regardless of the choice of $N_\mathrm{classes}$. To explore the uncertainty in the network's class assignments, we again employ the deep ensembles approach advocated by \cite{caldeira20} and, for each choice of $N_\mathrm{classes}$, train the network 100 times from different random initializations of the neuron weights.

\subsubsection{Network performance on a simulated data set}

\begin{figure}
\begin{center}
\includegraphics[width=0.45\textwidth]{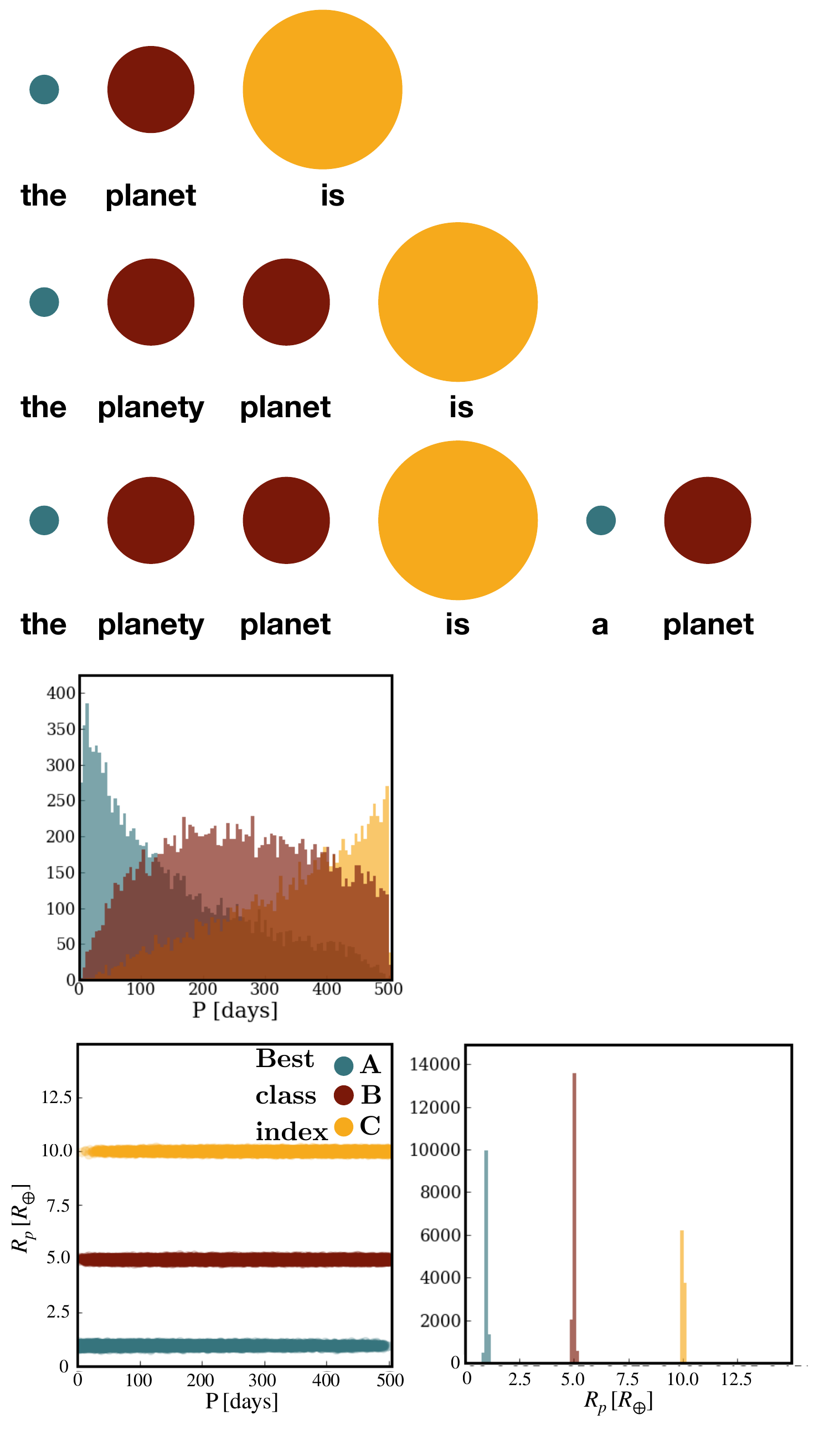}
\caption{Top: Allowed sentences in our toy planetary ``grammar." The three sizes of planet serve different grammatical functions in their systems, analogous to articles, adjectivey-nouns, and verbs. Bottom: The joint period-radius distribution of the planets in 10,000 test planetary systems generated according to these grammar rules. The planets are colourmapped by the context network's decision about their class membership---for this very artificial population, both the context network and the target network recover the truth exactly.}
\label{fig:simulatedClassifications}
\end{center}
\end{figure}

To test this setup, we simulate a data set of planets that follow an invented toy planetary grammar. A diagram of allowed planetary systems under this grammar is shown in the top panel of Figure~\ref{fig:simulatedClassifications}.

In the simulated data set, planets belong to three classes: small, or class $A$ ($R_p = 1 R_\oplus$); medium, or class $B$ ($R_p = 5 R_\oplus$); and large ($R_p = 10 R_\oplus$), or class $C$. The radius distributions are extremely narrow, and there is no overlap in radius between the classes (in other words, a planet's radius determines its class exactly). This is of course extremely unrealistic---the radius distribution of real planets is continuous, and planets do not fall neatly into discrete size categories---but useful in defining an unambiguous planet ``grammar'' (see below) and in testing the capabilities of the target and context networks. It will not be surprising if the target network, which has direct access to planet radius, is able to correctly identify the three planet categories; our hope is that the context network, which sees only contextual information but is evaluated on its agreement with the target network, will be able to identify the three categories too.

In our invented grammar, small planets act like articles (e.g. ``the" or ``a"), medium planets act like compoundable nouns (a single medium planet could represent the word ``planet," a sequence of two medium planets could represent the phrase ``planety planet," and so on), and large planets act like verbs (e.g. ``is"). 

Allowable planetary systems under this grammar include $[A,B,C]$ (``the planet is"), $[A,B,B,C]$ (``the planety planet is"), $[A,B,B,B,C]$ (``the planety planety planet is"), and $[A,B,B,C,A,B,B]$ (``the planety planet is a planety planet"), but \textit{not} $[A,A,A]$ (``the the the") or $[C, B, A, C]$ (``is planet the is"). 

To generate a simulated system, we first draw a random star from the catalog of \textit{Kepler} planet hosts satisfying $3 < \log{g \mathrm{[cm/s^2]}} < 5.3$, $2400 K < T_\mathrm{eff} < 9600 K$, and $R_* < 10R_\odot$. (These cuts are somewhat arbitrary, because our toy grammar does not depend on stellar properties.) We then draw a system multiplicity $m$ from a Zipfian distribution with maximum multiplicity $m_\mathrm{max} = 10$ and index $\beta=0.80$,\footnote{The one-sigma credibility interval of $\beta$ reported in \cite{sandford19} is $0.86^{+0.28}_{-0.29}$. In the course of preparing this manuscript, we uncovered a minor bug in the \texttt{Fortran} implementation of \texttt{forecaster} used to generate planetary systems in that paper, and re-ran the entire analysis. The results of the paper (in particular, the model comparison) did not change, but the median values of the posterior parameters shifted very slightly (while remaining $1\sigma$-consistent with the published values). The one-sigma credibility interval from the re-run is $\beta=0.80^{+0.28}_{-0.33}$, so here, we adopt $\beta=0.80$.} as found to be the best-fit model to the multiplicity distribution of \textit{Kepler} multi-planet systems in \cite{sandford19}:
\begin{equation}
\pdf(m) \propto
\begin{cases}
m^{-1-\beta_{ \mathrm{zipf} } } & \text{if } 1 \leq m \leq m_{\mathrm{max}} ,\\
0 & \text{otherwise},
\end{cases}
\end{equation}

We then draw periods for the $m$ planets in the system from a uniform distribution between 6.25 and 400 days. (Because class $A$ planets often begin systems, and class $C$ planets often end them, a planet's period contains weak information about its class membership---this is visible in the period histogram of Figure~\ref{fig:simulatedClassifications}.) We then populate the radii according to our grammar rules, and check for dynamical stability per the equations of \cite{fabrycky14}. If the system is stable, we keep it; otherwise, we discard it and draw new periods; if it is still unstable after 1000 period draws, we draw a new $m$ and start over. 

We generate a population of 10,000 training and 10,000 test systems according to this procedure, train the context and target networks on the training set, and then run the test set through the trained models. For both the training and test phases, we must also choose the number of categories $\mathrm{N}_{\mathrm{classes}}$ into which the networks attempt to sort the planets; when we choose the truth, $\mathrm{N}_{\mathrm{classes}} = 3$, both the target and context networks classify the simulated planets perfectly. When we choose $\mathrm{N}_{\mathrm{classes}} = 2$, the network divides class $B$ arbitrarily, lumping some $B$ planets in with the $A$ planets and some in with the $C$ planets. When we choose $\mathrm{N}_{\mathrm{classes}} > 3$, the networks begin to subdivide the true classes arbitrarily.

The lower panel of Figure~\ref{fig:simulatedClassifications} shows the radius and period distributions of the planets in the test set, colour-mapped according to the (perfectly accurate) class membership assigned by our \textit{context} network. The target network's classifications are identical, so we do not plot them separately.

Having demonstrated that the network is capable of classifying planets belonging to systems of this arbitrary grammar, we now turn to the real KOI data set to see what we can learn.

\subsection{Kepler systems, classified}\label{subsec:classified}

\begin{figure}
\begin{center}
\includegraphics[width=0.45\textwidth]{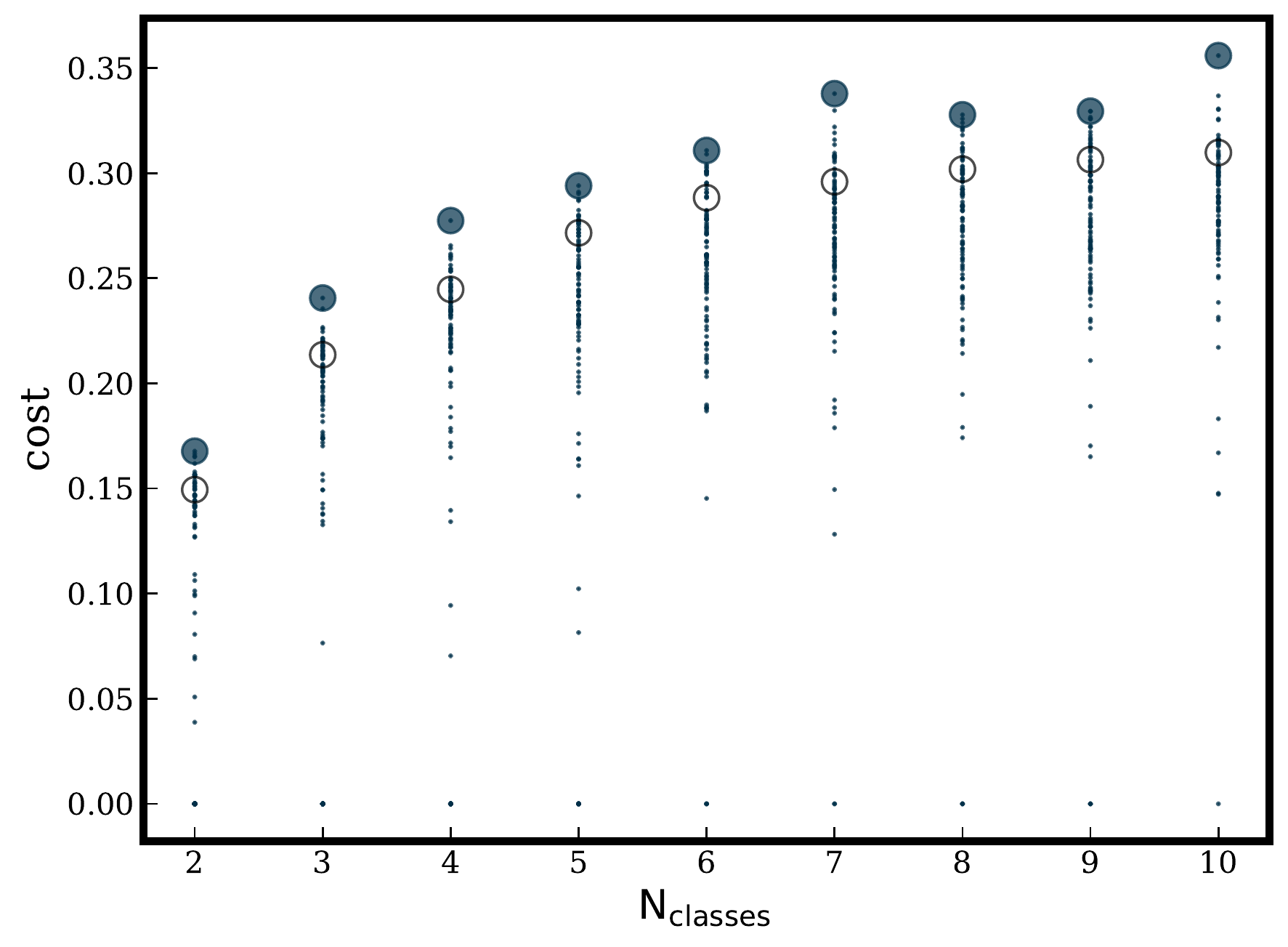}
\caption{The value of the cost function on the training set, after 500 training epochs, as a function of $\mathrm{N}_{\mathrm{classes}}$. For each choice of $\mathrm{N}_{\mathrm{classes}}$, we ran the network with 100 different random initializations of the network weights (small circles). The optimal value over the 100 trials is plotted as a large blue circle; the 20th-best value over the 100 trials is plotted as an open circle (see section~\ref{subsubsec:seed}, below). The optimal cost overall is (unsurprisingly) achieved when the model is at its most flexible, with $\mathrm{N}_{\mathrm{classes}} = 10$.}
\label{fig:cost_vs_N}
\end{center}
\end{figure}

We next apply the model shown in Figure~\ref{fig:classificationModel}, with the network architecture and hyperparameters given above, to the set of real KOI systems described in Section~\ref{sec:chap6_data}. For these real systems, we do not know in advance what the optimal choice of $\mathrm{N}_{\mathrm{classes}}$ is (if there is such a choice), so we run the network for $\mathrm{N}_{\mathrm{classes}} = 2$ to $10$, inclusive. For each choice of $\mathrm{N}_{\mathrm{classes}}$, we again do 100 training runs with different random initializations of the network weights, in order to explore the uncertainty of the network's class assignments. Figure~\ref{fig:cost_vs_N} shows the cost achieved on the training set as a function of $\mathrm{N}_{\mathrm{classes}}$. Each small circle represents the cost achieved by one of the 100 random seeds; the large circles represent the cost achieved by the best-performing network of the 100 at each choice of $N_\mathrm{classes}$.

In general, network performance improves with increasing $\mathrm{N}_{\mathrm{classes}} = 10$; the best-performing run overall (i.e., the one that maximizes $J$) is one where $\mathrm{N}_{\mathrm{classes}} = 10$. This is not surprising: when  $\mathrm{N}_{\mathrm{classes}}$ is large, the model is more flexible and can fit the training data better. The only exception to the general trend is the anomalously good performance of the best $\mathrm{N}_{\mathrm{classes}} = 7$ run, which is the second-best-performing run overall.

For all choices of $N_\mathrm{classes}$, there are some random seeds for which the networks experience a vanishing gradient problem and return a cost of 0.0, although the number of non-converging seeds decreases with increasing $N_\mathrm{classes}$. Thus, depending on random seed, there is a wide range of possible outcomes for the networks at each choice of $N_\mathrm{classes}$---this stands in stark contrast to the ensemble of 100 networks trained for the regression task (section~\ref{sec:chap6_regression}), which, regardless of random seed, made quite similar predictions on the test set (see the uncertainty bands plotted in yellow in Figure~\ref{fig:regressionResults}).

The goal of the next few paragraphs is to establish a way to interpret the network's classifications in light of this variation. To summarize, there are two nuisance parameters influencing the outcome of every training run of the classification networks: the choice of $N_\mathrm{classes}$, and the choice of random seed. We would like to identify classes or categorizations that are independent of these choices as far as possible.

\subsubsection{The choice of $N_\mathrm{classes}$}\label{subsubsec:nclasses}

\begin{figure*}
\begin{center}
\includegraphics[width=0.77\textwidth]{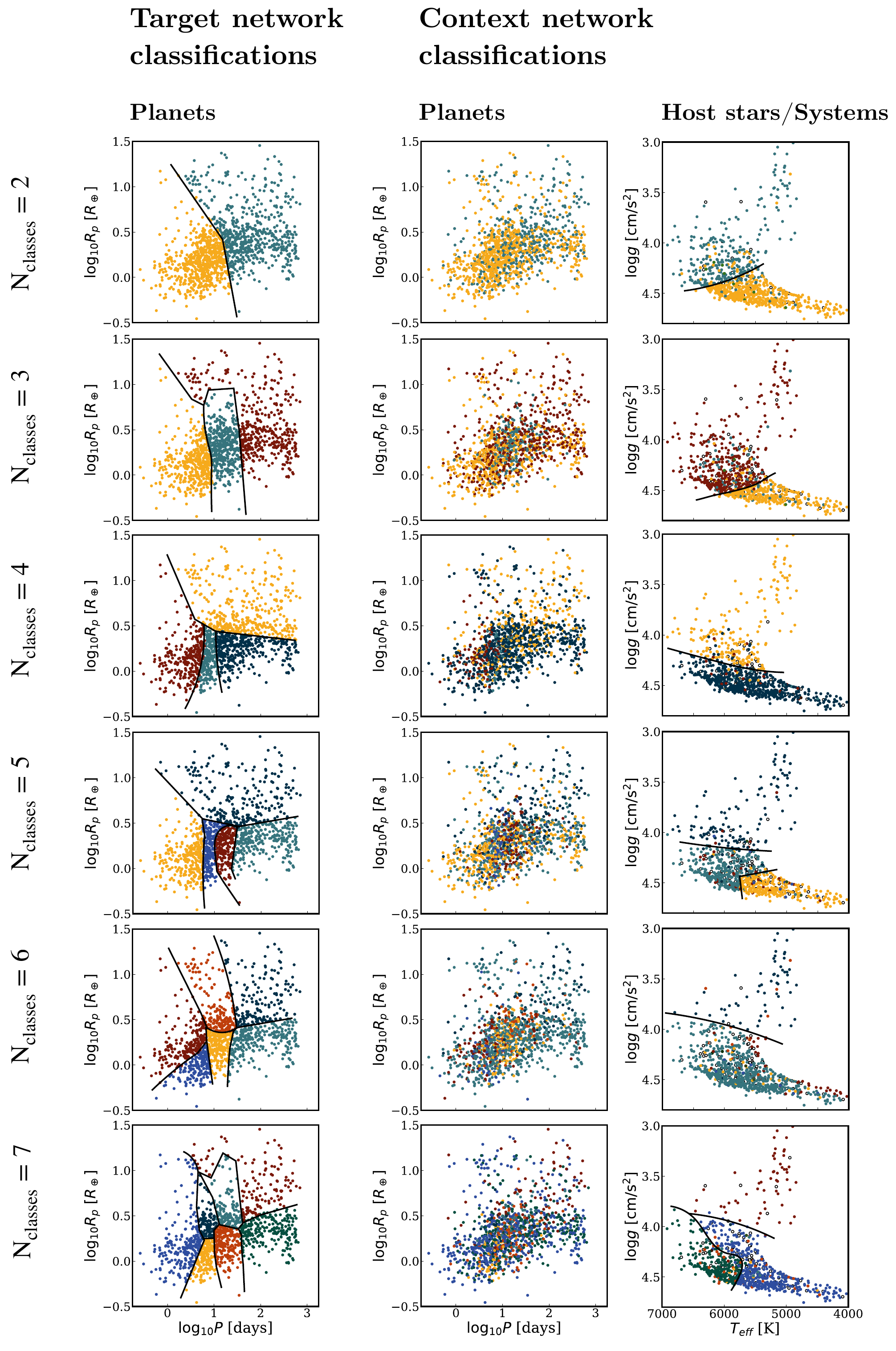}
\caption{Class assignments of the 1278 test-set planets by the \textit{target} network (left column) and by the \textit{context} network (middle and right columns), as a function of $\mathrm{N}_{\mathrm{classes}}$ (rows). By comparing the colour-mapping of points in the left and middle panels in each row, it is possible to evaluate the degree of agreement between the target and context network class assignments. Each data point in the left and middle columns represents one planet, while each data point in the right column represents one of the 984 host stars; host stars are colour-mapped by the modal class of the planets in their system. Host stars of systems with equal numbers of planets from multiple classes are plotted as open circles. Where class boundaries are obvious, we have drawn black lines as visual aids.}
\label{fig:realClassifications_nclasses}
\end{center}
\end{figure*}

In Figure~\ref{fig:realClassifications_nclasses}, we plot the network classifications of the 984 planetary systems (comprising 1278 planets) in the test set. We show the class assignments by the the optimally-seeded (i.e., best performing of the 100 random seeds at each choice of $N_\mathrm{classes}$) \textit{target} and \textit{context} networks for $\mathrm{N}_{\mathrm{classes}} = 2$ to $7$, to show how the networks' groupings change as the model grows more flexible. Each network outputs a vector of membership probabilities over all $\mathrm{N}_{\mathrm{classes}}$ classes for each test planet, so we choose to assign the planet to its highest-probability class and colour-map the data points in Figure~\ref{fig:realClassifications_nclasses} by class membership. (The colours assigned to the classes are arbitrary and not necessarily consistent from row to row in this figure, but are consistent \textit{within} rows.)

The left column shows the \textit{target} network's class boundaries in the $\log_{10}{R_P}$ vs. $\log_{10}{P}$ plane, and the middle and right columns show the \textit{context} network's class boundaries, in the $\log_{10}{R_P}$ vs. $\log_{10}{P}$ and $\log{g}$ vs. $T_{\mathrm{eff}}$ planes respectively. 

Visualizing this data is complicated for two reasons. First, the target and context class assignments for a particular planet do not always agree: as a result, while there is considerable overlap between e.g. the ``light blue'' class as picked up by the target network and the ``light blue'' class as picked up by the context network, these classes are not identical and do not have identical membership. The easiest way to evaluate the overlap (and consequently the degree of agreement between the target and context networks) is to compare the colour-mapping of planets between the left and middle panels in each row of Figure~\ref{fig:realClassifications_nclasses}.

Second, data points plotted in the $\log_{10}{R_P}$ vs. $\log_{10}{P}$ plane (left and middle columns) represent individual \textit{planets}, while data points plotted in the $\log{g}$ vs. $T_{\mathrm{eff}}$ plane (right column) represent \textit{host stars}, each of which corresponds to a system of one or more planets. We have chosen to colour-map each host star by the modal class of the planets in its system; for example, the host star of a system consisting of two ``dark blue'' planets and one ``orange'' planet would be plotted in dark blue. Host stars of systems with equal numbers of planets from multiple classes (e.g. a system of one ``dark blue'' planet and one ``orange'' planet) are plotted as open circles; these are a small minority of the stars overall.

What can we make of the two networks' class assignments? Recall that the \textit{target} network sees only $\log_{10}{P}$ and $\log_{10}{R_P}$ of the target planet, not the stellar features: this network therefore draws clean class boundaries in the radius vs. period plane and makes no distinctions along the $\log{g}$ and $T_{\mathrm{eff}}$ axes. 

The \textit{context} network sees $\log_{10}{P}$ and $\log_{10}{R_P}$ for the $2w$ neighbour planets, as well as $\log{g}$ and $T_{\mathrm{eff}}$ of the host star. Its class boundaries are cleanest in the $\log{g}$ vs. $T_{\mathrm{eff}}$ plane (right column), suggesting that the network relies heavily on information about the host star in order to decide on the target planet's class membership. This is not surprising because 80\% of the systems in the training set are single-planet systems, in which there are no neighbour planets, so the host star is all the context network sees.

However, the context network's class boundaries are also weakly visible in the radius vs. period plane (middle column), in the sense that planets from the same class tend to occupy similar regions of the parameter space, albeit with lots of overlap. Indeed, the context network's class assignments in the radius vs. period plane generally resemble a noisy version of the target network's class assignments. This makes sense because, although the context network cannot see the target planet's radius and period directly, it is rewarded for agreeing with the target network.

In the simplest version of the model, $\mathrm{N}_{\mathrm{classes}} = 2$, the networks draw simple boundaries. The target network divides planets into a short-period group ($\log_{10}{P} \lesssim 1$) and a long-period group, and the context network divides systems into those around dwarf stars ($\log{g} \gtrsim 4.4$) and those around giant stars. As the model complexity increases, these initial divisions broadly persist (although their exact positions shift), and new divisions are added. At $\mathrm{N}_{\mathrm{classes}} = 4$, the target network begins to distinguish (broadly) between small ($\log_{10}{R_p} \lesssim 0.5$) and large planets. At $\mathrm{N}_{\mathrm{classes}} = 5$, the context network begins to subdivide the dwarf host stars into cool ($T_{\mathrm{eff}} \lesssim 5500$ K) and warm groups. 

A simple way to identify persistent categories is to visualize the planets (or host stars) which belong to the same group as one another for \textit{every} choice of $\mathrm{N}_{\mathrm{classes}}$, up to and including $10$ (the most flexible model we tried). In other words, what groups of planets (or host stars) are never subdivided by the network? By our linguistic analogy, persistent planet groups would correspond to parts of speech, and persistent groups of stars---and the planetary systems they host---would correspond to different types of sentences.

\begin{figure*}
\begin{center}
\includegraphics[width=0.76\textwidth]{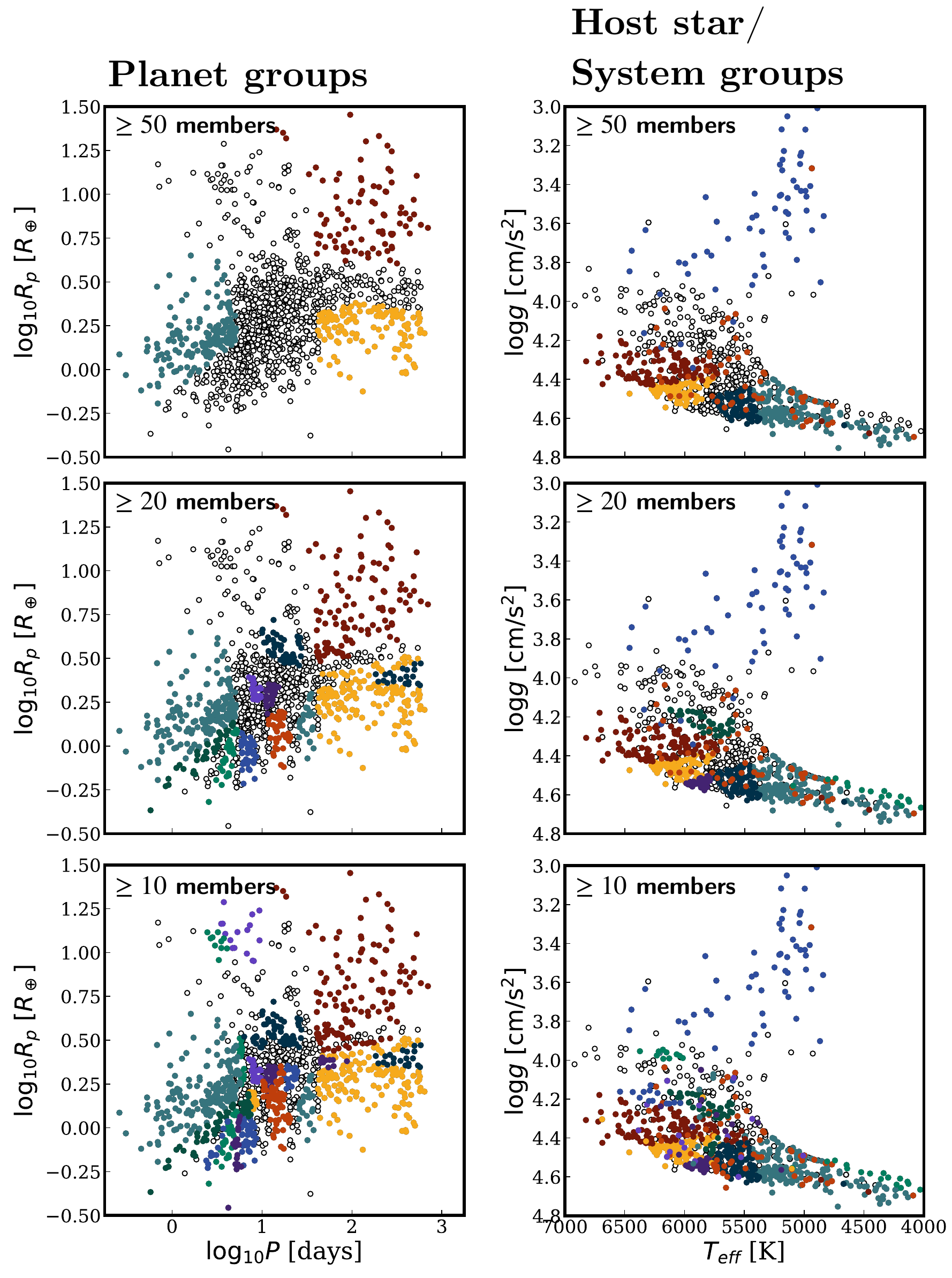}
\caption{How are the ``persistent'' groups constructed? Colourful points: Groups of planets (left column) and host stars (right column) which persist regardless of our choice of $\mathrm{N}_{\mathrm{classes}}$. The colour mapping of the groups is arbitrary and not consistent between the two columns; planets and stars which do not belong to the groups are plotted as open circles. Top row: Groups with $\geq 50$ members; there are 3 planet groups and 6 star groups of this size. Middle row: Groups with $\geq 20$ members (14 planet groups, 9 star groups). Bottom row: Groups with $\geq 10$ members (29 planet groups, 19 star groups).}
\label{fig:groupingBuildup}
\end{center}
\end{figure*}

\begin{figure*}
\begin{center}
\includegraphics[width=0.7\textwidth]{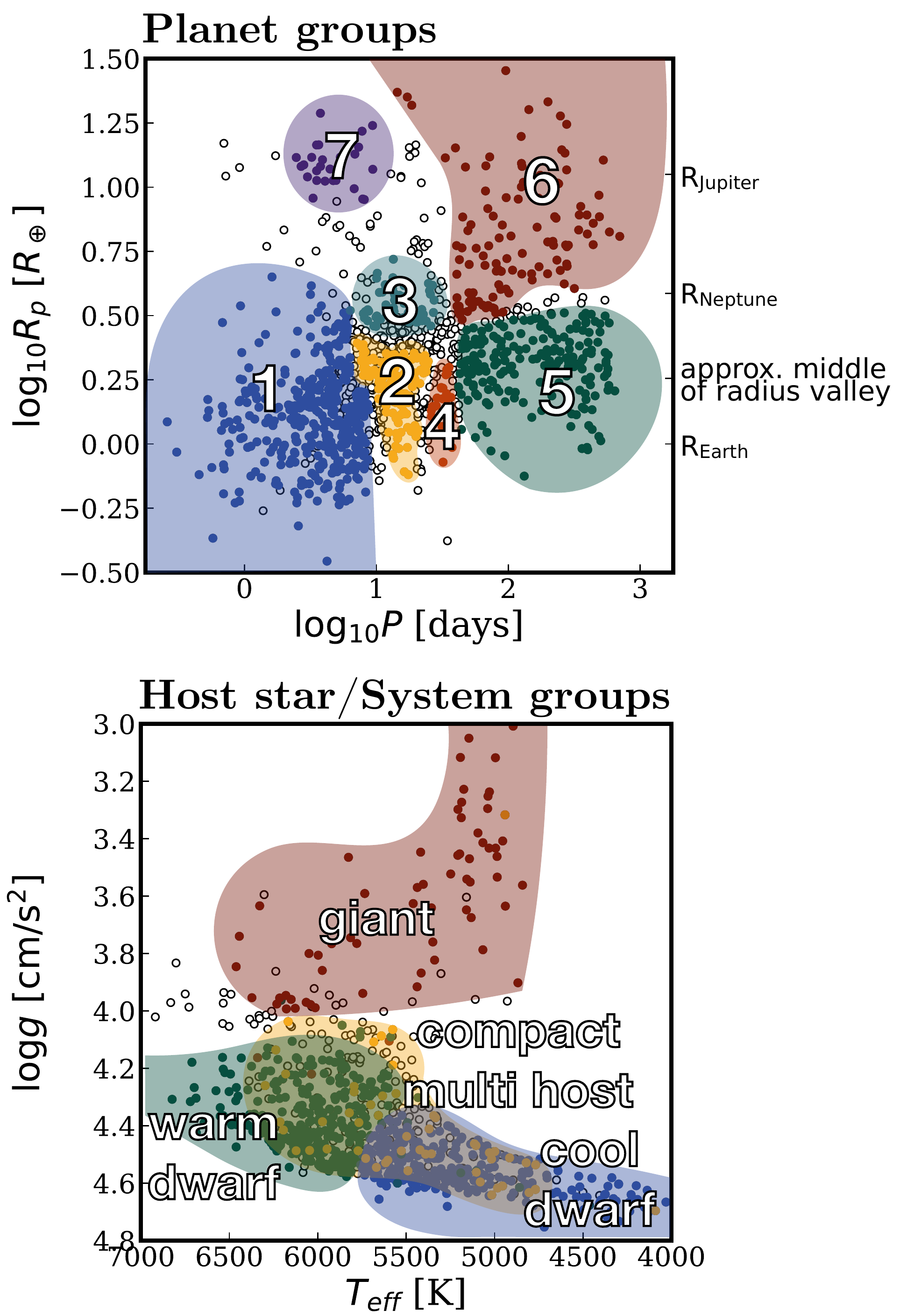}
\caption{Top panel: Seven groups of planets picked up by the target network. These are identified by merging the contiguous groups of $\geq 10$ members that persist for $\mathrm{N}_{\mathrm{classes}} = 2$ to $10$ from the bottom left panel of Figure~\ref{fig:groupingBuildup}. Bottom panel: Four groups of host stars picked up by the context network, identified by merging the corresponding groups in the bottom right panel of Figure~\ref{fig:groupingBuildup}.}
\label{fig:groupings}
\end{center}
\end{figure*}

\begin{figure*}
\begin{center}
\includegraphics[width=0.76\textwidth]{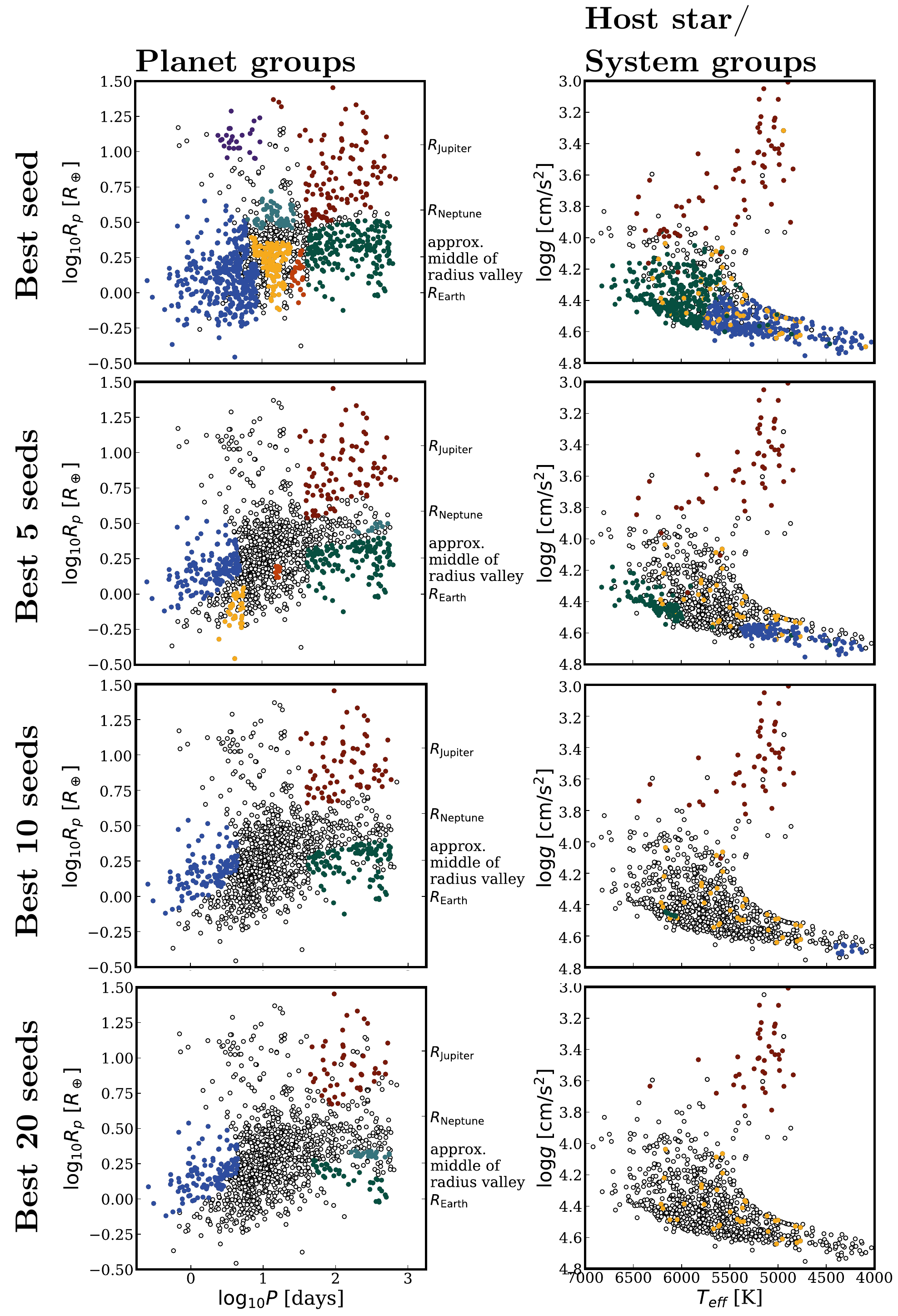}
\caption{How robust are the persistent groups to different random seed network initializations? Top row: The persistent planet (left) and stellar/system (right) groups agreed upon by the best-performing random network seed, over all choices of $\mathrm{N}_{\mathrm{classes}}$. (These are the same groups plotted in Figure~\ref{fig:groupings}.) Subsequent rows, moving downward: The groups agreed upon by the best 5, 10, and 20 seeds, again over all choices of $\mathrm{N}_{\mathrm{classes}}$.}
\label{fig:seedAgreement}
\end{center}
\end{figure*}

In Figure~\ref{fig:groupingBuildup}, we show these groups. In the top row, we plot all groups with 50 or more members. In the planetary plane, there are three such groups; in the stellar plane, there are six, although some of them are contiguous with one another. In the middle and bottom rows, we plot groups with $\geq 20$ and $\geq 10$ members, respectively. A pattern begins to emerge: smaller groups generally appear at the fringes of the large groups, rather than randomly among the unclassified points. By the bottom row, we can identify a few agglomerates built up of these smaller clusters. Meanwhile, the points in the lanes between these agglomerates are in ``contested territory,'' so to speak; they are lumped haphazardly with different neighbours for different choices of $\mathrm{N}_{\mathrm{classes}}$, so they do not belong to a persistent group of their own.

In the top panel of Figure~\ref{fig:groupings}, we plot the agglomerate groups that result when we merge the contiguous groups with $\geq 10$ members from the bottom left panel of Figure~\ref{fig:groupingBuildup}. These are the stable categories of planet in the $\log_{10}{R_P}$ vs. $\log_{10}{P}$ plane identified by our target network; there are seven such groups. We can interpret these groups (labeled by number in the top panel of Figure~\ref{fig:groupings}) as:

\setlist{noitemsep, topsep=0pt, parsep=0pt, partopsep=0pt, leftmargin=1.5cm, labelindent=0cm, labelwidth=\wd1, itemindent=*, labelsep=\dimexpr0.3cm-\wd1}

\renewcommand{\theenumi}{Group \arabic{enumi}.} 
\begin{enumerate}
    \item Hot sub-Neptunes ($P \lesssim 10$ days; $R_p \lesssim 4 R_\oplus$): 384 planets, plotted in blue.
    \item Short-period sub-Neptunes ($10 \lesssim P \lesssim 25$ days; $R_p \lesssim 2.5 R_\oplus$): 145 planets, plotted in yellow.
    \item Short-period Neptunes ($10 \lesssim P \lesssim 30$ days; $2.5 R_\oplus \lesssim R_p \lesssim 5.5 R_\oplus$): 59 planets, plotted in teal.
    \item Intermediate-period sub-Neptunes ($25 \lesssim P \lesssim 40$ days; $R_p \lesssim 2 R_\oplus$): 26 planets, plotted in orange.
    \item Long-period sub-Neptunes ($P \gtrsim 40$ days; $R_p \lesssim 3 R_\oplus$): 195 planets, plotted in green.
    \item Long-period giants ($P \gtrsim 40$ days; $R_p \gtrsim 3 R_\oplus$): 115 planets, plotted in red.
    \item Hot Jupiters ($2 \lesssim P \lesssim 10$ days; $R_p \gtrsim 10 R_\oplus$): 26 planets, plotted in purple.
\end{enumerate}

The remaining 328 planets are ``indeterminate'' and are plotted as open circles in the top panel of Figure~\ref{fig:groupings}.

As an aside, identifying these planetary groups explains the anomalously good performance of the model for $\mathrm{N}_{\mathrm{classes}} = 7$ (see Figure~\ref{fig:cost_vs_N}, above). If we compare the bottom left panel of Figure~\ref{fig:realClassifications_nclasses} to the top panel of Figure~\ref{fig:groupings}, we see that the class boundaries drawn by the $\mathrm{N}_{\mathrm{classes}} = 7$ target network approximate the boundaries between the stable groups in the planetary plane better than the class boundaries drawn by the target network for other choices of $\mathrm{N}_{\mathrm{classes}}$.

There are also four stable groups in the $\log{g}$ vs. $T_{\mathrm{eff}}$ plane, which we plot in the bottom panel of Figure~\ref{fig:groupings}. These are:
\vspace{0.1cm}

\setlist{noitemsep, topsep=0pt, parsep=0pt, partopsep=0pt, leftmargin=1cm, labelindent=0cm, labelwidth=\wd1, itemindent=*, labelsep=\dimexpr0.3cm-\wd1}

\renewcommand{\theenumi}{(\roman{enumi})} 
\begin{enumerate}
    \item Cool dwarf hosts ($T_\mathrm{eff} \lesssim 5700$ K; $\log{g} \gtrsim 4.3$): 329 stars (hosting 331 planets), plotted in blue.
    \item Warm dwarf hosts ($T_\mathrm{eff} \gtrsim 5700$ K; $\log{g} \gtrsim 4.1$): 330 stars (hosting 363 planets), plotted in green.
    \item Giant hosts ($\log{g} \lesssim 4.0$): 71 stars (hosting 75 planets), plotted in red.
    \item Hosts of compact multi-planet systems: 35 stars (hosting 76 planets), plotted in yellow. This group substantially overlaps the warm and cool dwarf groups in terms of $T_\mathrm{eff}$ and $\log{g}$ but is identified as a distinct cluster by the context network.
\end{enumerate}

The remaining 219 stars (hosting 433 planets) are ``indeterminate'' and are plotted as open circles in the bottom panel of Figure~\ref{fig:groupings}. 

We identify the cool dwarf, warm dwarf, and giant host star groups by merging the contiguous groups of $\geq 10$ members from the bottom right panel of Figure~\ref{fig:groupingBuildup}. The fourth group, of compact multi-planet system hosts, is different---unlike the other three groups, membership is determined primarily by the character of the planetary system around the host star, not by the host star's location in the $\log{g}$ vs. $T_{\mathrm{eff}}$ plane. The indeterminate host stars likewise overlap the other groups in this plane.

\subsubsection{Robustness of the groups over the ensemble of differently-seeded networks}\label{subsubsec:seed}

In section~\ref{subsubsec:nclasses} above, we considered only the best-performing network out of the 100 that we ran (from different random seeds) at each choice of  $\mathrm{N}_{\mathrm{classes}}$. Now, we consider: how far do the groups we identified above persist across different choices of random seed? In other words, are these groups robust to perturbations to the network model? Answering this question will provide some insight into the ``uncertainty'' of the groupings, which is more difficult to quantify for the classification task than for the regression task.

To answer this question, we repeat the procedure described in section~\ref{subsubsec:nclasses} three more times, to build three analogues to the set of agglomerate groups plotted in Figure~\ref{fig:groupings}. This time, we insist that the groups are not only never subdivided at any choice of $\mathrm{N}_{\mathrm{classes}}$, but also never subdivided across the best 5, 10, and 20 random seeds, respectively. The resulting groups are plotted in the second, third, and fourth rows of Figure~\ref{fig:seedAgreement}. The choice to stop at 20 is somewhat arbitrary, but as we consider versions of the network with less and less optimal performance, we expect less and less agreement with the optimal network on the basis of poorness-of-fit to the training set (Figure~\ref{fig:cost_vs_N} shows the cost achieved by both the best and the 20th-best network at each $\mathrm{N}_{\mathrm{classes}}$); if we were to include all 100 seeds, we would expect no agreement---and no persistent groups---at all.

Moving downward through the rows of Figure~\ref{fig:seedAgreement}, we can see that, as we insist upon the agreement of more and more of the differently-seeded networks, fewer groups persist, and those that do are generally smaller than their Figure~\ref{fig:groupings} counterparts, sliced away at the edges.

In the planet plane (left column of Figure~\ref{fig:seedAgreement}), the groups agreed upon by all 20 network are a subset of the hot sub-Neptunes (Group 1), two subsets of the long-period sub-Neptunes (Group 5), and a subset of the long-period giants (Group 6). It is sensible that these three groups are most robust: they are large clusters at the edges of the radius vs. period plane, as far as possible from the contested territory in the middle.

Similarly, in the stellar plane (right column of Figure~\ref{fig:seedAgreement}, the warm dwarf system and cool dwarf system groups gradually erode away to nothing as we insist on agreement between more and more networks. However, the giant system group only erodes slightly, and the compact multi-planet system group is almost unchanged.

Broadly, we interpret the more robust groups---those which are agreed upon by more of the randomly seeded networks---to be more ``real,'' in the sense that they are less dependent on any specific configuration of network weights. Correspondingly, we place more confidence in the compact multi-planet system group and the giant star host group than in the warm and cool dwarf host groups, and more confidence in the hot sub-Neptunes, long-period sub-Neptunes, and long-period giants than in the other planet groups.

In what follows, we nevertheless consider all of the groups identified in Figure~\ref{fig:groupings}, for two reasons: first, to explore any differences between the cool and warm dwarf systems and the overlapping ``indeterminate'' systems, and second, because the colour-mapping of the seven planet groups makes the following figures more legible.

\subsection{System grammar}\label{subsec:grammar}

Having identified these persistent planet and star groups and explored their robustness, we next ask: Is there an identifiable, physically interpretable ``grammar'' of planetary systems? What patterns can we observe in the ordering of planets within and between systems? 

\begin{figure}
\begin{center}
\includegraphics[width=0.45\textwidth]{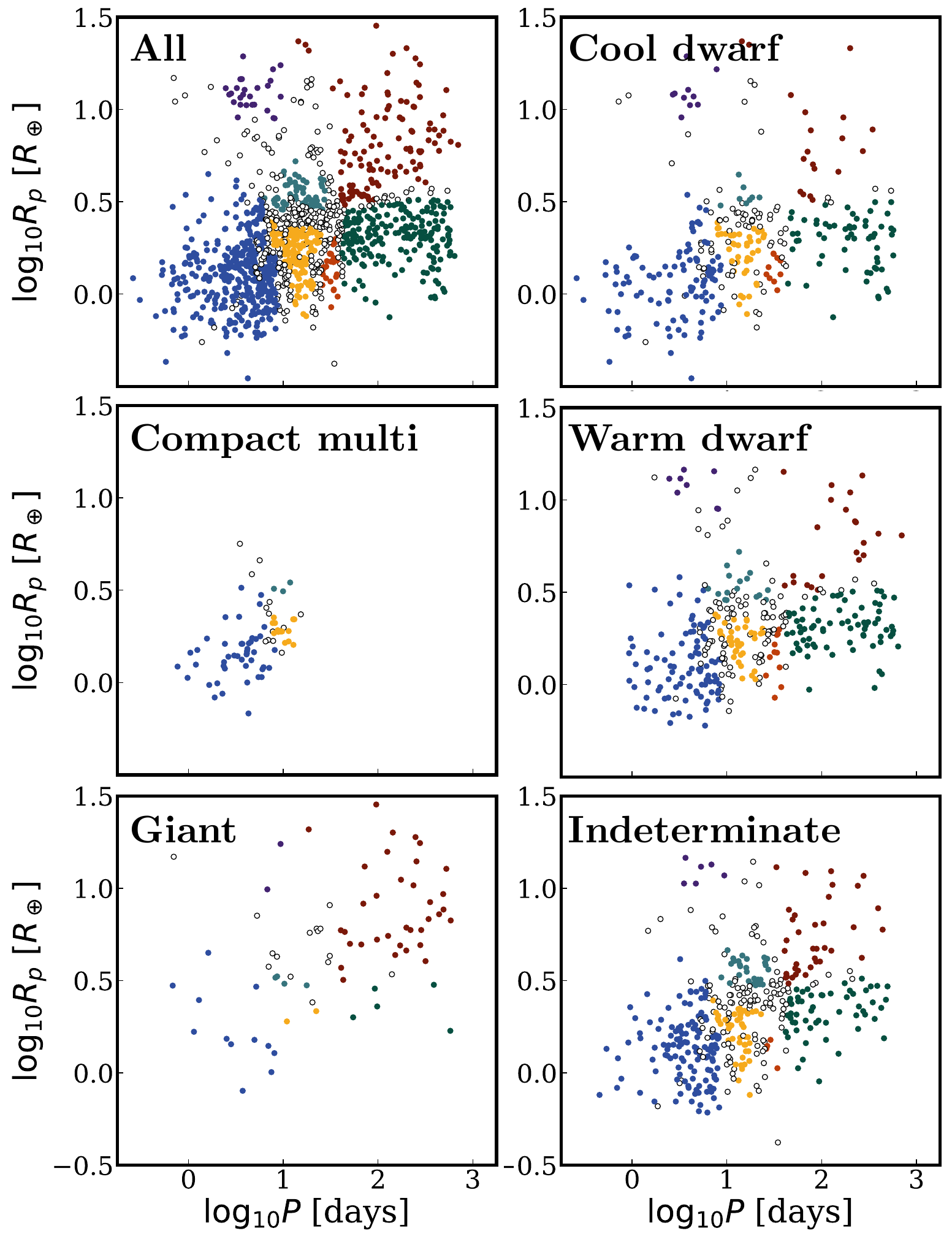}
\caption{Planetary systems from the test set, sorted by what stellar group the host star belongs to.}
\label{fig:sysType}
\end{center}
\end{figure}

In Figure~\ref{fig:sysType}, we separate out the test set systems into the four host star groups identified above, plus systems orbiting indeterminate stars, and show which planets orbit each type of host. From this figure, we can see that the stellar and planetary groupings discussed above correspond to one another. In other words, there are certain categories of planet which are more likely to be found around certain categories of host star. This correspondence is strongest in the compact multi-planet systems and the giant systems, the two host star groups we found to be most robust (see section~\ref{subsubsec:seed} above).

Most notably, the compact multi-planet systems contain only planets from Group 1 (hot sub-Neptunes; blue), Group 2 (short-period sub-Neptunes; yellow), and Group 3 (short-period Neptunes; teal), plus a handful of indeterminate planets. Planets from Groups 1 and 2 are, moreover, overrepresented in the compact multi-planet systems: 12\% of Group 1 planets and 10\% of Group 2 planets belong to compact multis, compared to only 6\% of planets overall. There are no planets from Groups 4 to 7 in these systems.

Meanwhile, planets from Group 6 (long-period giants; red) are likely to be found around giant host stars. Quantitatively, although only 6\% of planets overall orbit giant stars, 30\% of Group 6 planets orbit giant stars. This is once again the result of the transit method's selection bias: only giant planets are big enough to be observable in transit around bright giant stars.

There are also trends between planet group membership and system multiplicity. In particular, planets from Group 5 (long-period sub-Neptunes; green), Group 6 (long-period giants; red), and Group 7 (hot Jupiters; purple) are overrepresented among the single-planet systems. Overall, only 62\% of planets are single; meanwhile, 78\% of Group 5 planets, 70\% of Group 6 planets, and 96\% of Group 7 planets are single. Meanwhile, short- (but not ultra-short-) period planets are overrepresented in two-planet systems: 29\% of Group 2 planets (short-period sub-Neptunes; yellow) and 34\% of Group 3 planets (short-period Neptunes; teal) belong to two-planet systems, compared to only 20\% of planets overall.

\begin{figure}
\begin{center}
\includegraphics[width=0.33\textwidth]{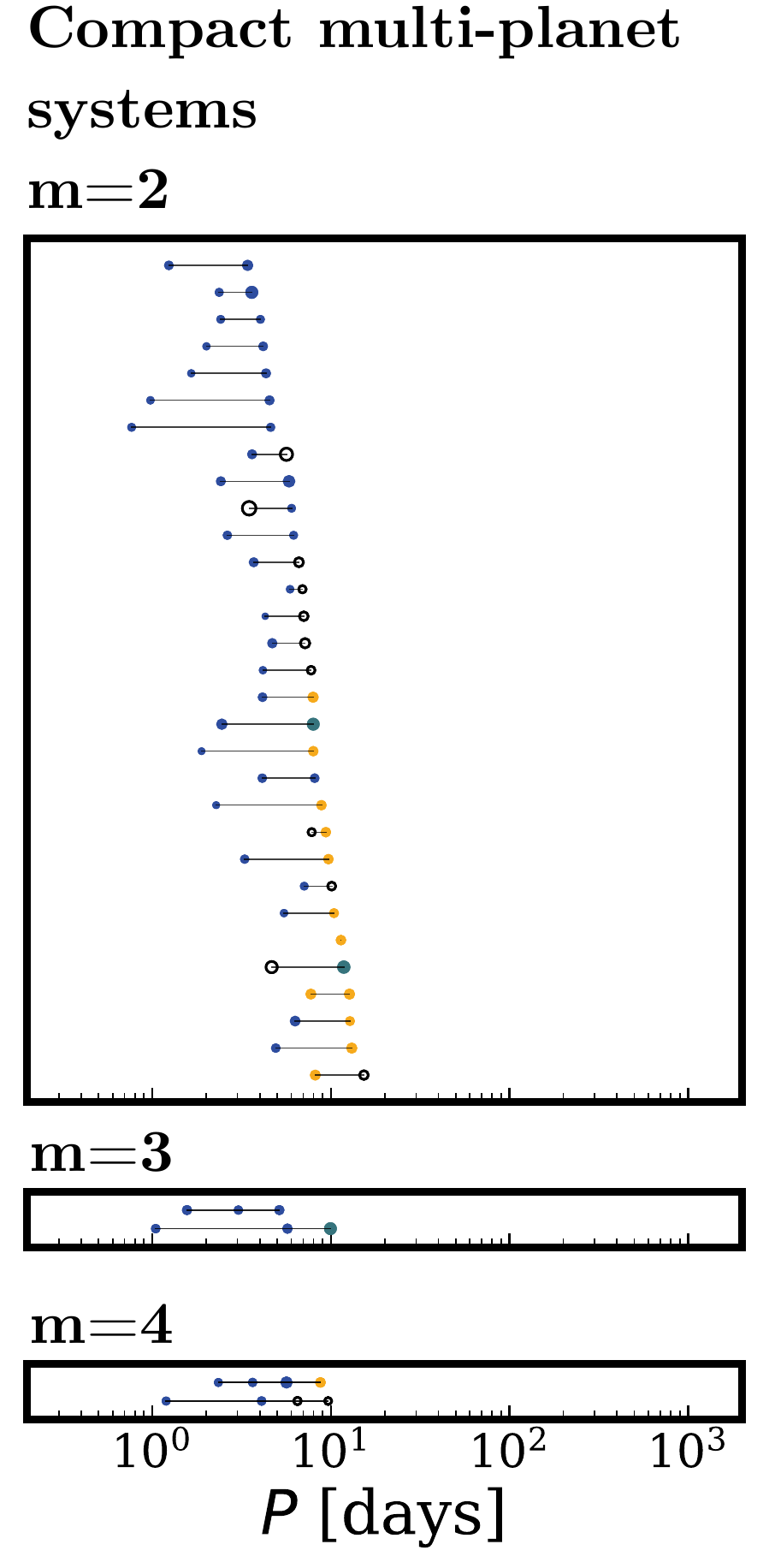}
\caption{The 35 compact multi-planet systems (comprising 76 planets), sorted into panels by multiplicity $m$. Each row represents one system; each circle represents one planet, with area proportional to $R_p$. Systems are sorted by outermost planet period. Planets are colour-coded by their planet group membership (see top panel of Figure~\ref{fig:groupings}.)}
\label{fig:compactMulti}
\end{center}
\end{figure}

\begin{figure}
\begin{center}
\includegraphics[width=0.31\textwidth]{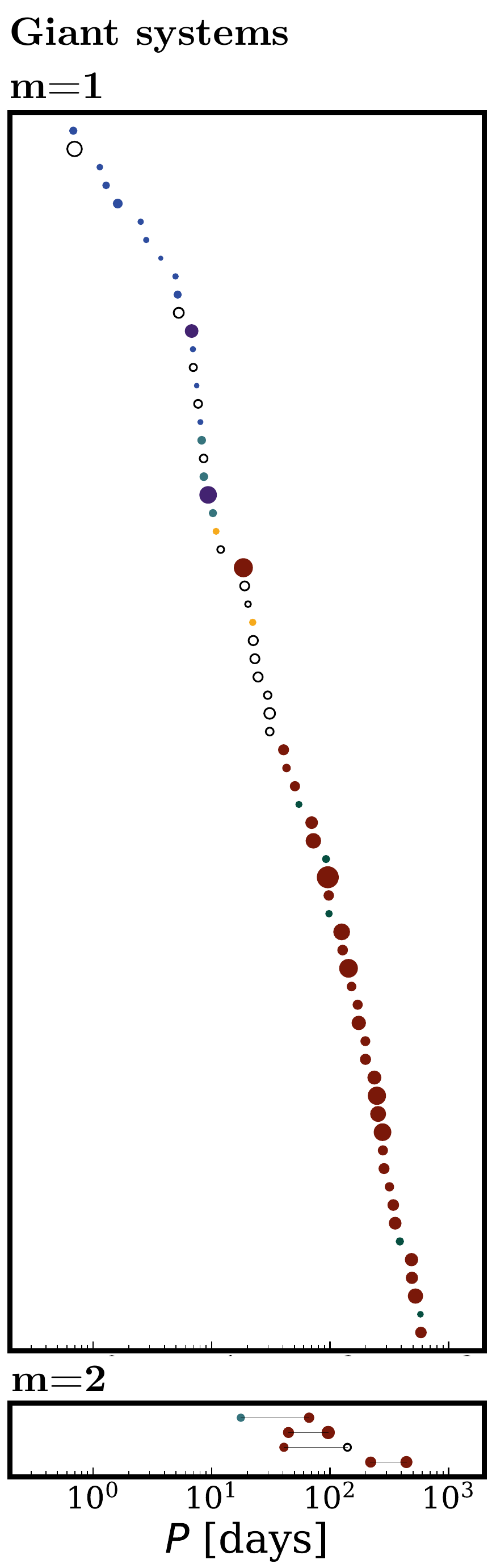}
\caption{The 71 giant systems (comprising 75 planets), sorted into panels by multiplicity $m$. Each row represents one system; each circle represents one planet, with area proportional to $R_p$. Systems are sorted by outermost planet period. Planets are colour-coded by their planet group membership (see top panel of Figure~\ref{fig:groupings}.)}
\label{fig:giant}
\end{center}
\end{figure}

\begin{figure*}
\begin{center}
\includegraphics[width=0.89\textwidth]{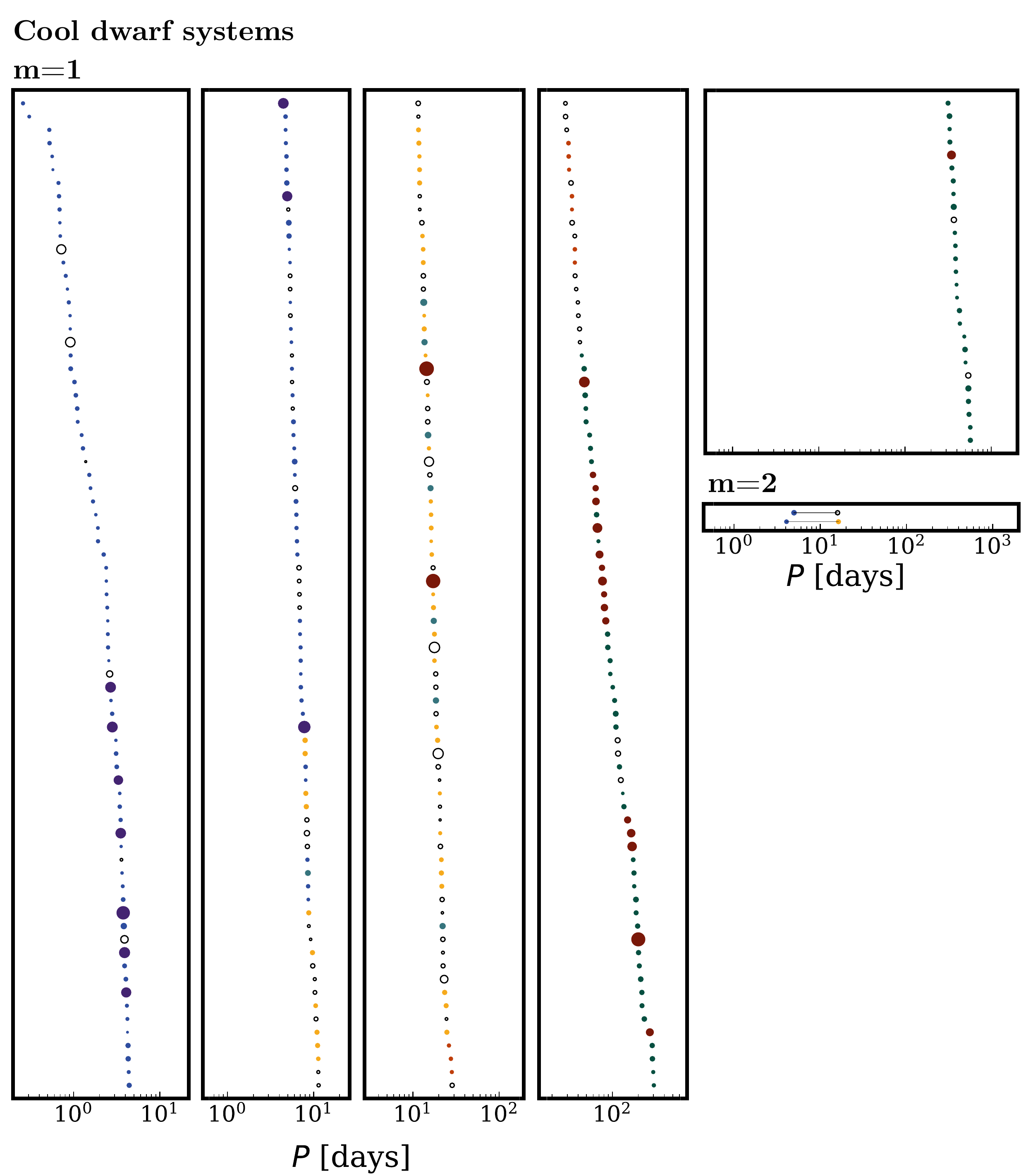}
\caption{The 329 planetary systems (331 planets) orbiting cool dwarf stars, sorted into panels by multiplicity $m$. Each row represents one system; each circle represents one planet, with area proportional to $R_p$. Systems are sorted by outermost planet period (note the change in x-axis limits across the 1-planet systems). Planets are colour-coded by their planet group membership (see top panel of Figure~\ref{fig:groupings}.)}
\label{fig:coolDwarf}
\end{center}
\end{figure*}

\begin{figure*}
\begin{center}
\includegraphics[width=0.92\textwidth]{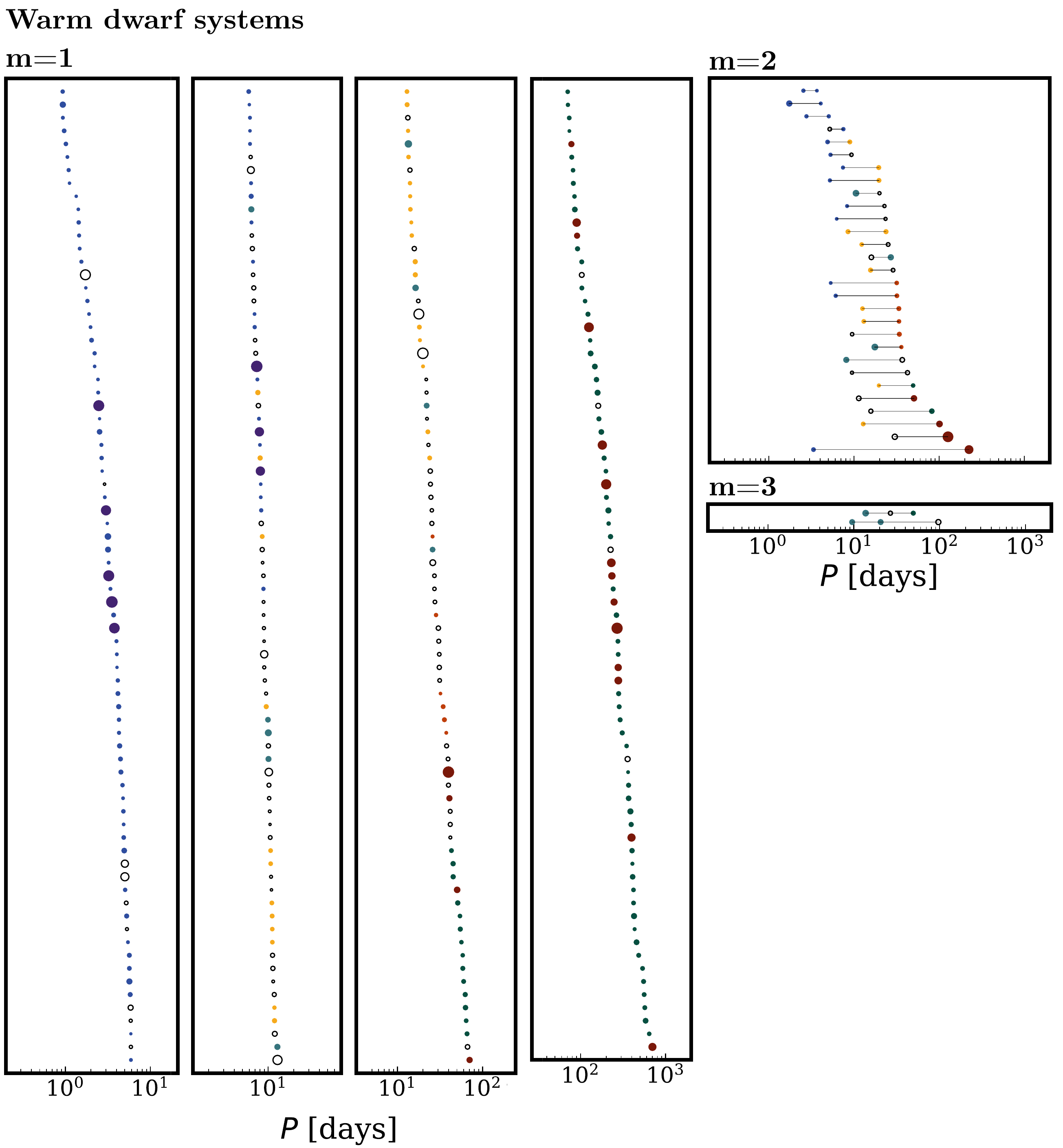}
\caption{The 330 planetary systems (363 planets) orbiting warm dwarf stars, sorted into panels by multiplicity $m$. Each row represents one system; each circle represents one planet, with area proportional to $R_p$. Systems are sorted by outermost planet period (note the change in x-axis limits across the 1-planet systems). Planets are colour-coded by their planet group membership (see top panel of Figure~\ref{fig:groupings}.)}
\label{fig:warmDwarf}
\end{center}
\end{figure*}

\begin{figure*}
\begin{center}
\includegraphics[width=\textwidth]{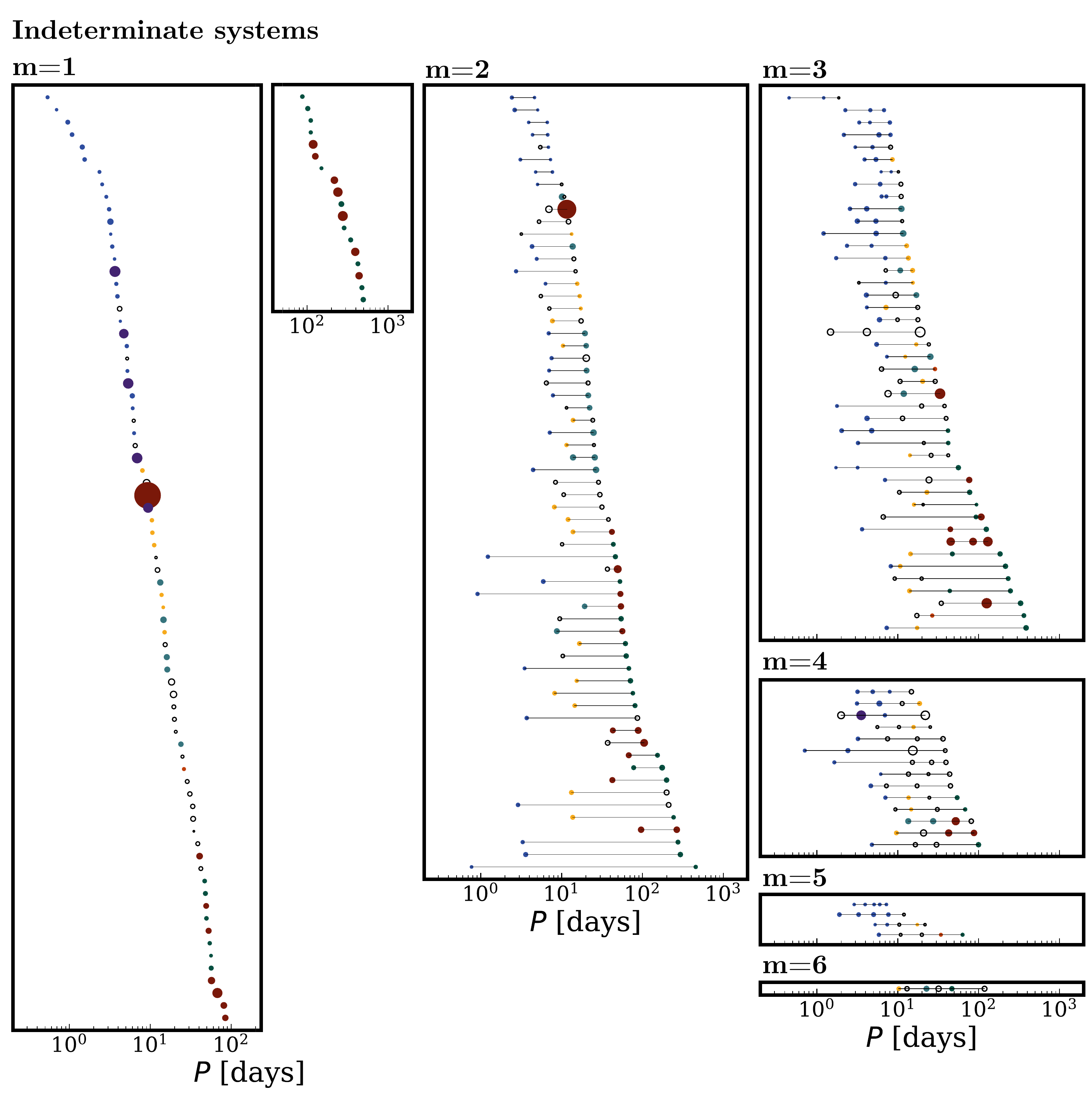}
\caption{The 219 planetary systems (433 planets) orbiting indeterminate host stars, sorted into panels by multiplicity $m$. Each row represents one system; each circle represents one planet, with area proportional to $R_p$. Systems are sorted by outermost planet period (note the change in x-axis limits across the 1-planet systems). Planets are colour-coded by their planet group membership (see top panel of Figure~\ref{fig:groupings}.)}
\label{fig:indeterminate}
\end{center}
\end{figure*}

In Figures~\ref{fig:compactMulti}, \ref{fig:giant}, \ref{fig:coolDwarf}, \ref{fig:warmDwarf}, and~\ref{fig:indeterminate}, we plot every planetary system from the test set orbiting a compact multi host, giant host, cool dwarf host, warm dwarf host, or indeterminate host star, respectively. In each figure, each row represents one planetary system, and systems are sorted by increasing outermost planet period. Each dot represents one planet, size-scaled according to the planetary radius and colour-coded by its planet group (1-7, or indeterminate) from the top panel of Figure~\ref{fig:groupings}.

In all five of these figures, certain trivial patterns arise immediately from the period ordering of systems in the panels: blue (Group 1) planets, which come from the shortest-period group, occur more frequently toward the top of each multiplicity panel, and green (Group 5) and red (Group 6) planets, from the long-period groups, occur more frequently toward the bottom. Within systems, which are individually sorted by period, certain sequences are impossible given the period ranges of the groups: a green planet (Group 5) cannot precede a yellow (Group 2), for example, because Group 5 falls to the right of Group 2 in the period-radius plane.

Also evident are the well-known patterns which result from the observational biases of the transit method. In general, planet size and period are correlated, i.e. planets at longer periods are larger; this pattern is sculpted by \textit{Kepler}'s increasing incompleteness toward the lower-right quadrant of the period-radius plane (see e.g. \citealt{petigura:2013}, figure 1). 

Within the multi-planet systems ($m \geq 2$) of all five figures, an interesting pattern arises which perhaps sheds some light on why the planet group boundaries from the top panel of Figure~\ref{fig:groupings} fall where they do: it is rare for neighbouring planets in a system to come from non-contiguous classes in the map (i.e., classes that do not share a border). For example, it is rare for a green planet (Group 5) to immediately follow a yellow (Group 2), or for an orange (Group 4) to immediately follow a blue (Group 1). Only $10\%$ of neighbouring planet pairs (28/294) come from non-contiguous classes. 

Vastly more common are either (1) successive planets from the same class, e.g. ``blue - blue'' or (2) neighbouring planets from neighbouring groups. $42\%$ (123/294) of neighbouring planet pairs come from either the same group or contiguous groups, and the remaining $49\%$ (143/294) of neighbouring planet pairs include at least one indeterminate planet.

The commonness of same-group pairs and contiguous pairs is overall superficially consistent with a ``peas in a pod'' picture of planetary systems, where planets are correlated in size with their neighbours (see e.g. \citealt{weiss:2018}).

We now investigate the specific characteristics of planetary systems orbiting each type of star. In the below, it is important to distinguish between e.g. ``cool-dwarf-group host stars,'' which are specific stars belonging to the cool dwarf group in the lower panel of Figure~\ref{fig:groupings}, and host stars which are cool dwarfs in terms of temperature and surface gravity. Some of the latter belong to the compact multi-host group, or the indeterminate group, instead of the ``cool dwarf group.''

\subsubsection{Compact multi-planet system grammar}

In Figure~\ref{fig:compactMulti}, we show the 35 compact multi-planet systems (comprising 76 planets) from the test set. Host stars in this group do not form a distinct cluster in the $T_\mathrm{eff}$ vs. $\log{g}$ plane, but rather overlap both the warm dwarf and cool dwarf groups: clearly, in order to cluster these systems together, the context network has picked up on information encoded in the constituent planets of these systems rather than information in the host stars alone. And indeed, the planetary systems belonging to this group are more homogeneous than the systems belonging to any other group.

As noted above, planets from Groups 4 to 7 are completely absent from this group. Meanwhile, small, short-period planets are overrepresented, with 12\% of planets from Group 1 (hot sub-Neptunes; blue) and 10\% of planets from Group 2 (short-period sub-Neptunes; yellow) belonging to compact multis, compared to only 6\% of planets overall.

There are no single-planet systems in this group; all systems have between 2 and 4 planets. The average system multiplicity is highest in this group, with an average of 2.2 planets per system. The average period ratio between neighbouring planets in compact multi-planet systems is 1.8, which is smaller than for any other group except the warm dwarf systems. 

Compact multi-planet systems therefore have the smallest, shortest-period planets, with the highest system multiplicity, of any group. 

\subsubsection{Giant system grammar}

In Figure~\ref{fig:giant}, we show the 71 planetary systems (comprising 75 planets) from the test set orbiting stars from the giant group ($\log{g} \lesssim 4.0$). 

As a consequence of the observational biases of the transit method (see above), intermediate-to-long-period giant planets are heavily overrepresented in this group: 30\% of Group 6 planets (long-period giants; red) orbit giant group stars, compared to only 6\% of planets overall. In contrast, sub-Neptune-sized planets are underrepresented across the entire period range: in order of increasing period, only 3\% of Group 1 planets (blue), 1\% of Group 2 planets (yellow), 0\% of Group 4 planets (orange), and 3\% of Group 5 planets (green) orbit host stars in this group.

Multi-planet systems are underrepresented around giant group stars.  7\% of the planetary systems overall, but only 2\% of the multi-planet systems, belong to this group. No systems with 3 or more planets orbit giant group stars. 

Giant group stars therefore preferentially host large, lonely planets. 

\subsubsection{Cool dwarf system grammar}

In Figure~\ref{fig:coolDwarf}, we show the 329 planetary systems (comprising 331 planets) from the test set orbiting stars from the cool dwarf group ($T_\mathrm{eff} \lesssim 5700$ K; $\log{g} \gtrsim 4.3$).

Which planet groups are overrepresented in cool dwarf group systems? 38\% of Group 4 planets (intermediate-period sub-Neptunes; orange), 31\% of Group 5 planets (long-period sub-Neptunes; green), and 38\% of Group 7 (hot Jupiters; purple) orbit cool-dwarf-group hosts, compared to only 26\% of planets overall.

Other groups are underrepresented: only 14\% of Group 3 planets (short-period Neptunes; teal) and 17\% of Group 6 (long-period giants; red) belong to cool dwarf group systems.

Meanwhile, planets from Group 1 (hot sub-Neptunes; blue) and Group 2 (short-period sub-Neptunes; yellow), as well as the indeterminate planet group (open circles), are represented in proportion to their occurrence in the test set overall.

Multi-planet systems are also heavily underrepresented around cool dwarf group stars:  33\% of the planetary systems in the test set orbit giant stars, but only 1\% of the multi-planet systems (and zero $m > 3$ systems) do. The cool dwarf systems have the lowest average multiplicity of any group, with only 1.01 planets per system.

In summary, the cool dwarf systems have a higher proportion of small long-period planets than the population of test set planets overall. A large fraction of the hot Jupiters orbit cool dwarf group hosts, but few long-period giants do. This group includes very few multi-planet systems, and no systems with three or more planets.

\subsubsection{Warm dwarf system grammar}

In Figure~\ref{fig:warmDwarf}, we show the 330 planetary systems (comprising 363 planets) from the test set orbiting stars from the warm dwarf group ($T_\mathrm{eff} \gtrsim 5700$ K; $\log{g} \gtrsim 4.1$). (This group overlaps the Sun, which has $T_{\mathrm{eff}} = 5780$ K and $\log{g} = 4.4$.) 

Small, intermediate-to-long period planets are overrepresented around warm dwarf group stars. 46\% of Group 4 planets (intermediate-period sub-Neptunes; orange) and 37\% of Group 5 planets (long-period sub-Neptunes; green) orbit warm dwarf group stars, compared to only 28\% of planets overall. Hot Jupiters are also modestly overrepresented, with 31\% of Group 7 planets (purple) orbiting warm dwarf group stars.

Meanwhile, long-period giants are underrepresented, as only 17\% of Group 6 planets (red) orbit warm dwarf group stars.

Multi-planet systems are also underrepresented in the warm dwarf group: only 16\% of multi-planet systems orbit warm dwarf group stars, compared to 34\% of test set systems overall. The average multiplicity of warm dwarf systems is 1.1. Multi-planet systems in this group are more tightly spaced than in any other group; the average period ratio between neighbouring planets is 1.5.

Overall, the trends in the warm dwarf group are the same as those in the cool dwarf group, but slightly weaker. In particular, the underrepresentation of multi-planet systems is much less extreme than in the cool dwarf group. This group also contains the most closely spaced planetary systems, on average.

\subsubsection{Indeterminate systems}

In Figure~\ref{fig:indeterminate}, we show the 219 planetary systems (comprising 433 planets) orbiting indeterminate host stars. This group has more planets than any other.

Unsurprisingly, indeterminate-type planets are overrepresented in this group: 39\% of indeterminate planets, compared to 34\% of planets overall, orbit indeterminate stars. Group 3 planets (short-period Neptunes; teal) are also overrepresented: 47\% of these planets orbit indeterminate stars.

Some groups are underrepresented as well: only 15\% of Group 4 planets (intermediate-period sub-Neptunes; orange), 29\% of Group 5 planets (long-period sub-Neptunes; green), and 23\% of Group 7 planets (hot Jupiters; purple) orbit indeterminate hosts. These trends are the opposite of the trends among the cool and warm dwarf group systems, where Groups 4, 5, and 7 were overrepresented and Group 3 were underrepresented. 

Multi-planet systems are strongly overrepresented among the indeterminate stars. 64\% of multi-planet systems, compared to only 22\% of systems overall, orbit indeterminate stars. The average multiplicity of the indeterminate systems is 2.0, which is only slightly smaller than the average multiplicity in the compact multi-planet system group. 

All of the highest-multiplicity systems in the test set---the four 5-planet systems, and the lone 6-planet system---belong to the indeterminate group. The indeterminate group systems are also the most spaced out on average: the mean period ratio between neighbouring planets is 4.9, which is higher than any other group.

Overall, the indeterminate systems are eclectic. This is not surprising, given that these are the systems that explicitly did not belong to any of the large clusters picked out by the context network.

\section{Discussion \& Conclusions}
\label{sec:chap6_discussion}
Here, we have explored two avenues to understanding  planetary systems as ordered sequences, in which the arrangement of individual planets contains information beyond that contained in the planets themselves. In other words, we have explored ways to understand planets in the context of their systems---their host star, their sibling planets, and their position among them. We specifically investigate a data set consisting of 4286 \textit{Kepler} Objects of Interest, grouped into 3277 planetary systems of multiplicity ranging from 1 to 7.

We first explore a regression problem: is it possible to accurately predict the radius and period of a planet, if the radii and periods of its surrounding planets are known? We find that our trained model indeed can. We compare our network's predictions to those of a naive model, which accounts only for basic orbital stability, and find that the mean absolute error (MAE) of the network's planetary radius predictions is a factor of 2.1 smaller than the MAE of the naive model's radius predictions. Similarly, the MAE of the network's period predictions is a factor of 2.1 smaller than the MAE of the naive model's period predictions.

We find furthermore that the network's predictions improve with increasing system multiplicity, and that neither the network nor the naive model are able to make meaningful predictions about most single-planet systems, indicating that the network relies on information about a planet's neighbours, not its host star, to make its predictions. The only exception to this is a small subset of single giant planets orbiting giant stars: the network picks up on the association between planet radius and stellar surface gravity that results from the observational bias of the transit method (i.e., the bias that only giant planets have deep enough transits to be detectable around bright giant stars) and uses it to accurately predict high radii for this subset of planets.

We next explore a planetary classification model based on recent advances in unsupervised linguistic part-of-speech tagging. Our model consists of two networks; one which sees the ``target" planet whose properties we are trying to predict, and the other which sees the surrounding stellar and planetary ``context.'' The network trains by maximizing the mutual information between the class assignments of the target network and those of the context network, a principle which has proved very successful in linguistic applications but has not previously been used in exoplanet science. 

We find that the target network tends to group certain planets together, regardless of the number of classes with which the model is told to label the data. These groups correspond to parts of speech by our linguistic analogy. The seven planet groups (moving roughly counterclockwise around the radius vs. period plane) are: Group 1, hot sub-Neptunes ($P \lesssim 10$ days; $R_p \lesssim 4 R_\oplus$); Group 2, short-period sub-Neptunes ($10 \lesssim P \lesssim 25$ days; $R_p \lesssim 2.5 R_\oplus$); Group 3, short-period Neptunes ($10 \lesssim P \lesssim 30$ days; $2.5 R_\oplus \lesssim R_p \lesssim 5.5 R_\oplus$); Group 4, intermediate-period sub-Neptunes ($25 \lesssim P \lesssim 40$ days; $R_p \lesssim 2 R_\oplus$); Group 5, long-period sub-Neptunes  ($P \gtrsim 40$ days; $R_p \lesssim 3 R_\oplus$); Group 6, long-period giants ($P \gtrsim 40$ days; $R_p \gtrsim 3 R_\oplus$); and Group 7, hot Jupiters ($2 \lesssim P \lesssim 10$ days; $R_p \gtrsim 10 R_\oplus$). Of these groups, the hot sub-Neptunes (Group 1), long-period sub-Neptunes (Group 5), and long-period giants (Group 6) are most robust, i.e. most agreed-upon by the ensemble of the 20 best-performing randomly seeded training runs.

Meanwhile, the \textit{context} network tends to group certain host stars (and by extension, planetary systems) together: hosts of compact multi-planet systems; giant hosts ($\log{g} \lesssim 4.0$); cool dwarf hosts ($T_\mathrm{eff} \lesssim 5700$ K; $\log{g} \gtrsim 4.3$); and warm dwarf hosts ($T_\mathrm{eff} \gtrsim 5700$ K; $\log{g} \gtrsim 4.1$). The compact multis are distinguished by the character of their planetary systems, not the character of their host stars: stars in this category substantially overlap both the cool and warm dwarf groups in the $\log{g}$ vs. $T_\mathrm{eff}$ plane. By our linguistic analogy, these four system groups correspond to different types of sentences. The compact multi-planet systems and the giant systems are substantially more robust (i.e., more agreed-upon by the ensemble of 20 best-performing randomly seeded training runs) than the cool and warm dwarf groups.

We can identify certain ``grammatical patterns'' in the planetary systems from each of the four system groups. The compact multi-planet systems, which are the most homogeneous of any group, contain exclusively planets with periods less than 11 days and radii of less than $5.7 R_\oplus$. These systems have the highest average multiplicity of any group, and the second-lowest average period ratio, after the warm dwarf group.

Stars in the giant group preferentially host single-planet systems. Intermediate-to-long period giant planets are dramatically overrepresented, and sub-Neptune-sized planets dramatically underrepresented, in this group, an effect which is at least partially explained by the observational biases of the transit method. 

Systems orbiting stars in the cool dwarf group are also overwhelmingly single-planet systems. Intermediate- and long-period sub-Neptunes are overrepresented in this group, as are hot Jupiters, but short-period Neptunes and long-period giants are underrepresented. 

Systems orbiting stars in the warm dwarf group show similar trends in planet representation, but have slightly higher average multiplicity. They also have the smallest average period ratio between neighbouring planet pairs---in other words, the multi-planet systems of the warm dwarf group are more tightly spaced than the multis in other groups.

The remaining ``indeterminate'' systems are an eclectic mix of systems not easily included in any of the other groups. The planets which are overrepresented in the cool and warm dwarf groups tend to be underrepresented in the indeterminate group, and vice versa. The highest-multiplicity systems in the test set (including all of the 5-planet systems, and the lone 6-planet system) belong to this group; consequently, this group has the second-highest average multiplicity after the compact multi group. This group also has the least-tightly-spaced multi-planet systems of any group, with the largest mean period ratio between neighbouring planets. To resolve the indeterminate group into more meaningful categories, we will need substantially more multi-planet systems than are presently known.

Overall, these methods show promise for picking up on subtle patterns in the arrangement of planetary systems. We caution that our data set is only one window onto the exoplanet population, so the conclusions offered here---particularly the groupings of planets and systems---are tentative. In future, it would be very interesting to expand our sample to include e.g. TESS systems, where cool stars are much better represented than among the \textit{Kepler} target stars. It would be extremely interesting also to apply these methods to a data set of transiting planets containing more hot Jupiters, which are overall larger, closer-in, and lonelier than the planets investigated here. Prospects for inducing planetary grammar can only improve as we gather more data.

\section*{Acknowledgments}

The authors thank the referee for a thorough and careful report. ES, DK, \& MC acknowledge support from the Columbia University Data Science
Institute ``Seed Funds Program''. Thanks to members of the Cool Worlds Lab, the Cambridge exoplanets group, and Zephyr Penoyre for useful discussions in preparing this manuscript. 

\section*{Data Availability Statement}

The data underlying this article were accessed from the NASA Exoplanet Archive's Cumulative KOI Data Table at \url{https://exoplanetarchive.ipac.caltech.edu/cgi-bin/TblView/nph-tblView?app=ExoTbls&config=cumulative} on 30 September 2020. The derived data generated in this research will be shared on reasonable request to the corresponding author.

\bibliographystyle{mnras}
\bibliography{bib}
\bsp

\appendix
\section{Analytic constraints on missing planets}
\label{sec:appendix}
Here, we consider the analytic constraints we can place on the period and radius, respectively, of a hypothetical target planet, based on (1) the assumption that the planetary system is dynamically stable and (2) knowledge of the periods and radii of the planets adjacent to the target planet. According to \cite{fabrycky14}, dynamical stability demands that the separation $\Delta$ between neighbouring planets in units of mutual Hill radii satisfies:
\begin{equation}
    \Delta > 2\sqrt{3},
\end{equation}

where $\Delta$ is defined as

\begin{equation}
    \Delta \equiv \frac{a_\mathrm{outer} - a_\mathrm{inner}}{R_H}
\end{equation}

and $R_H$ by

\begin{equation}
    R_H \equiv \left(\frac{M_\mathrm{inner} + M_\mathrm{outer}}{3M_*}\right)^{1/3}\frac{a_\mathrm{inner} + a_\mathrm{outer}}{2}.
\end{equation}

\cite{fabrycky14} further suggest a conservative heuristic stability criterion for neighbouring pairs of planets, but we do not consider it here.

The goal of this section is to adopt the most conservative possible constraints (in the sense of adopting the fewest further assumptions about the planetary system), yielding the widest physically allowable intervals of $R_{p,\mathrm{target}}$ and $P_\mathrm{target}$. 

\subsubsection{Period}\label{subsubsec:period}

Assuming circular orbits, we can bracket the period of any unobserved target planet by the periods of its inner and outer neighbour planets: $P_\mathrm{inner} < P_\mathrm{target} < P_\mathrm{outer}$. (Circularity is a conservative assumption, given that orbital eccentricity of either neighbour planet would impose stricter bounds on the allowed period range of the middle one.) If a target planet is the innermost in its system, we can similarly bracket its period by

\begin{equation}
    \sqrt{\frac{4\pi^2 (R_*+R_p)^3}{G(M_*+M_p)}} < P_\mathrm{target} < P_\mathrm{outer},
\end{equation}

where the inner limit comes from calculating the period of an orbit at a semi-major axis of $R_* + R_p$ using Kepler's Third Law.

If the planet is the outermost in its system, we can bracket its period by $P_\mathrm{inner} < P_\mathrm{target} \leq P_\mathrm{max\,in\,data\,set}$, where the latter is 1694 days.

\subsubsection{Radius}\label{subsubsec:radius}

The radius constraint is slightly more involved, and requires us to adopt a mass-radius relationship for the KOIs in our sample. In brief, the procedure for constraining $R_{p,\mathrm{target}}$ is:
\begin{enumerate}
    \item Using the mass-radius relationship, calculate $M_{p,\mathrm{inner}}$ and $M_{p,\mathrm{outer}}$ from $R_{p,\mathrm{inner}}$ and $R_{p,\mathrm{outer}}$;
    \item Use these to calculate the maximum allowable stable mass of the target planet, $M_{p,\mathrm{target,max}}$;
    \item Again use the mass-radius relationship to translate this maximum mass into an upper limit on radius, $R_{p,\mathrm{target,max}}$.
\end{enumerate}

Step 2 involves a rearrangement of the equations of \cite{fabrycky14} (see above). For the planetary system to be dynamically stable, it must hold both that:
\begin{equation}
    M_{p,\mathrm{target}} < \frac{M_*}{\sqrt{3}}\left(\frac{a_\mathrm{target} - a_\mathrm{inner}}{a_\mathrm{target} + a_\mathrm{inner}}\right)^{-3} - M_{p,\mathrm{inner}}
\label{eq:innerConstraint}
\end{equation}
and 
\begin{equation}
    M_{p,\mathrm{target}} < \frac{M_*}{\sqrt{3}}\left(\frac{a_\mathrm{outer} - a_\mathrm{target}}{a_\mathrm{outer} + a_\mathrm{target}}\right)^{-3} - M_{p,\mathrm{outer}}.
\label{eq:outerConstraint}
\end{equation}

Since $M_*$ is known, and $a$ and $M_p$ are known for both neighbours, these constraints represent upper limits on $M_{p,\mathrm{target}}$ as a function of $a_\mathrm{target}$, which (in accordance with our circularity assumption, above) could take any value between $a_\mathrm{inner}$ and $a_\mathrm{outer}$. 

The maximum allowable mass for the target planet \textit{overall}, i.e. anywhere in the allowable range of semi-major axis, is at the intersection of the two constraints, where $a_\mathrm{target}$ satisfies
\begin{equation}
    \left(\frac{a_\mathrm{target} - a_\mathrm{inner}}{a_\mathrm{target} + a_\mathrm{inner}}\right)^{3} - \frac{M_{p,\mathrm{inner}}\sqrt{3}}{M_*} = \left(\frac{a_\mathrm{outer} - a_\mathrm{target}}{a_\mathrm{outer} + a_\mathrm{target}}\right)^{3} - \frac{M_{p,\mathrm{outer}}\sqrt{3}}{M_*}.
\end{equation}

We adopt this maximum allowable mass, $M_{p,\mathrm{target,max}}$, as our upper mass constraint, and translate it to an $R_{p,\mathrm{target,max}}$ with our assumed mass-radius relationship. 

\begin{figure}
\begin{center}
\includegraphics[width=0.45\textwidth]{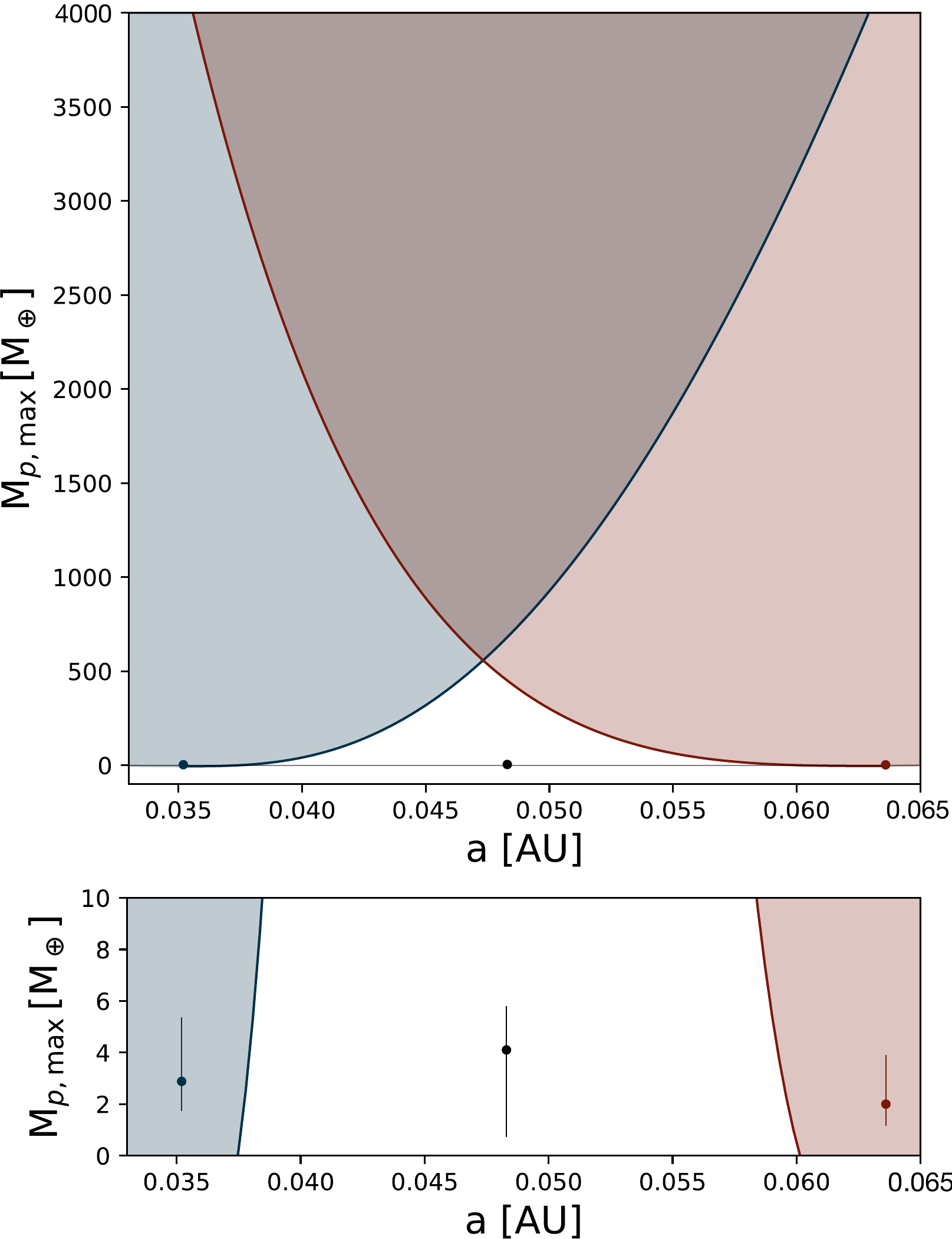}
\caption{An investigation into which regions of parameter space are excluded by simple analytic stability criteria for representative planet Kepler-1073c, which is bounded by inner planet KOI 2055.03 and outer planet KOI 2055.04. Top panel: Masses above the blue line are excluded by the inner planet; masses above the red line are excluded by the outer planet. Lower panel: A zoomed-in view, showing the Forecaster-predicted $1\sigma$ mass ranges for all three planets based on their NEA-reported $1\sigma$ radius ranges. In practice, the analytic upper mass limits are generally much too large to meaningfully constrain the middle planet's mass. }
\label{fig:kepler1073c}
\end{center}
\end{figure}

For the innermost planet in any system, we may similarly derive an $M_{p,\mathrm{target,max}}$ based on the outer-planet mass constraint, Equation~\ref{eq:outerConstraint}. We must also make sure that the planet is small enough not to be tidally destroyed by the star, so we can use the mass-radius relationship to translate the upper limit on $M_{p,\mathrm{target}}$ as a function of $a_\mathrm{target}$ into an upper limit on $R_{p,\mathrm{target}}$ as a function of $a_\mathrm{target}$, then calculate the star's Roche limit $d$ as a function of that $R_{p,\mathrm{target}}$:
\begin{equation}
    d = R_{p,\mathrm{target}} \left(2\frac{M_*}{M_{p,\mathrm{target}} }\right)^{1/3}.
\end{equation}

We take $M_{p,\mathrm{target,max}}$ to be the maximum allowed $M_{p,\mathrm{target}}$ at which $a_\mathrm{target}$ exceeds $d$.

For the outermost planet, we can adopt the largest allowable mass by the inner-planet mass constraint (Equation~\ref{eq:innerConstraint}) at the semi-major axis of the maximum period in the data set,
\begin{equation}
a_\mathrm{max} = \left(\frac{GM_*P_\mathrm{max\,in\,data\,set}^2}{4\pi^2}\right)^{1/3}.
\end{equation}

In Figure~\ref{fig:kepler1073c}, we take as an example the first three planets of the Kepler-1073 (KOI 2055) system, and plot the mass upper limits we derive for the middle planet (Kepler-1073c) based on its inner neighbour (KOI 2055.03) and its outer neighbour (KOI 2055.04). The mass constraints from the inner and outer neighbour as a function of $a_\mathrm{target}$ are plotted as blue and red curves, respectively, with the corresponding excluded regions shaded in blue and red. The true mass and semi-major axis of the middle planet are plotted as a black point. Here, we have adopted the probabilistic broken power law mass-radius relationship of \cite{chen:2017}, implemented in the \texttt{forecaster} package. We use \texttt{forecaster} to translate the NEA-reported $1\sigma$ uncertainty range of $R_p$ into a corresponding range of $M_p$ for the inner and outer planets. 

This system is representative of our results for the KOI data set overall: in practice, the upper limit on $M_{p,\mathrm{target}}$ that we derive by this procedure---and the corresponding upper limit on $R_{p,\mathrm{target}}$---is orders of magnitude larger than the true value, and indeed orders of magnitude too large to be practically relevant for our prediction problem.


\label{lastpage}
\end{document}